\documentclass[aps,floats,twocolumn,prd,showpacs,nofootinbib]{revtex4}

\usepackage[dvips]{graphicx} % 
\usepackage{epsfig,amsmath}
\usepackage{amssymb}

\usepackage{rotate}
\usepackage{color}

\DeclareFontFamily{OT1}{pzc}{}
\DeclareFontShape{OT1}{pzc}{m}{it}%
            {<-> s * [1.10] pzcmi7t}{}
\DeclareMathAlphabet{\mathscr}{OT1}{pzc}%
                                {m}{it}
\definecolor{RedWine}{rgb}{0.743,0,0}
\definecolor{RoyalBlue}{rgb}{0.25,.41,.88}
\definecolor{ForestGreen}{rgb}{.13,.54,.13}

\newcommand{\cmnt}[1]{}

\newcommand{\mnras}{MNRAS}

\newcommand{\apjl}{Astrophys. J. Lett.}

\def\widecheck#1{\check{#1}}
\def\bm#1{\textbf{\em #1}}
\def\fnl{f_\mathrm{NL}}

\def\Q{\mathcal{Q}}
\def\mmu{\mathcal{M}}

\def\bt{b_e}
\def\btp{b_{ep}}
\def\cH{\mathcal{H}}

\def\ln{{\rm ln}}

\def\vk{\mathrm{\bf k}}

\def\vx{\mathrm{\bf x}}

\def\fnl{f_{\rm NL}}

\def\apjl{Astrophys.\ J.\ Lett.}
\def\mnras{Mon.\ Not.\ R.\ Astron.\ Soc.}

\def\apj{Astrophys.\ J.}

\def\apjs{Astrophys.\ J. Supp.}

\def\prd{Phys.\ Rev.\ D}

\def\jcap{{JCAP\ }}

\newcommand{\be}{\begin{equation}}
\newcommand{\ee}{\end{equation}}
\newcommand{\bea}{\begin{eqnarray}}
\newcommand{\eea}{\end{eqnarray}}
\def\ba#1\ea{\begin{align}#1\end{align}}
\newcommand{\D}{\Delta}
\newcommand{\rhob}{\bar{\rho}_m}
\newcommand{\Om}{\Omega_m}

\newcommand{\s}{\sigma}
\renewcommand{\d}{\delta}
\renewcommand{\a}{\alpha}

\newcommand{\eps}{\varepsilon}

\renewcommand{\v}[1]{\bm{#1}}
\newcommand{\<}{\langle}
\renewcommand{\>}{\rangle}

\renewcommand{\k}{\kappa}

\newcommand{\zt}{\tilde{z}}
\newcommand{\chib}{\bar{\chi}}
\newcommand{\chit}{\tilde{\chi}}

\newcommand{\iMpch}{\,h/{\rm Mpc}}

\def\M{\mathcal{M}}

\newcommand{\refeq}[1]{Eq.~(\ref{eq:#1})}          
\newcommand{\refeqs}[2]{Eqs.~(\ref{eq:#1})--(\ref{eq:#2})}

\newcommand{\reffig}[1]{Fig.~\ref{fig:#1}}          

\newcommand{\refsec}[1]{Sec.~\ref{sec:#1}}
\newcommand{\refapp}[1]{App.~\ref{app:#1}}
\newcommand{\vs}{\nonumber\\} 

\def\nhatt{\hat{\tilde{n}}}
\def\vnhatt{\hat{\tilde{\v{n}}}}

\begin{document}

\title{
Large-scale clustering of galaxies in general relativity
}

\author{Donghui Jeong}
\affiliation{Theoretical Astrophysics,
	California Institute of Technology, 
	Mail Code 350-17, Pasadena, CA  91125, USA}
\author{Fabian Schmidt}
\affiliation{Theoretical Astrophysics,
	California Institute of Technology, 
	Mail Code 350-17, Pasadena, CA  91125, USA}
\author{Christopher M. Hirata}
\affiliation{Theoretical Astrophysics,
	California Institute of Technology, 
	Mail Code 350-17, Pasadena, CA  91125, USA}

\begin{abstract}
Several recent studies have shown how to properly
calculate the observed clustering of galaxies in a relativistic context,
and uncovered corrections to the Newtonian calculation that become
significant on scales near the horizon.  
Here, we retrace these calculations and 
show that, on scales approaching 
the horizon, the observed galaxy power spectrum 
depends strongly on which gauge is assumed to 
relate the intrinsic fluctuations in galaxy density to matter perturbations
through a linear bias relation.  
Starting from simple physical assumptions, we derive a gauge-invariant 
expression relating galaxy density perturbations to matter density
perturbations on large scales, and show that it reduces to 
a linear bias relation in synchronous-comoving gauge, corroborating an 
assumption made in several recent papers.  We evaluate the resulting observed 
galaxy power spectrum, and show that it leads to corrections similar
to an effective non-Gaussian bias corresponding to a local 
$f_{\rm NL,eff} \lesssim 0.5$.  
This number can serve 
as a guideline as to which surveys need to take into account relativistic 
effects.  We also discuss the scale-dependent bias induced by primordial
non-Gaussianity in the relativistic context, which again is simplest
in synchronous-comoving gauge.   
\end{abstract}
\date{27 July 2011}


\maketitle

%%%%%%%%%%%%%%%%%%%%%%%%%%%%%%%%%%%%%%%%%%%%%%%%%%%%%%%%%%%%%%%%%%%%%%%%%%%%%
%%%%%%%%%%%%%%%%%%%%%%%%%%%%%%%%%%%%%%%%%%%%%%%%%%%%%%%%%%%%%%%%%%%%%%%%%%%%%
%%%%%%%%%%%%%%%%%%%%%%%%%%%%%%%%%%%%%%%%%%%%%%%%%%%%%%%%%%%%%%%%%%%%%%%%%%%%%
\section{Introduction}
\label{sec:intro}

The clustering of galaxies and other large-scale structure (LSS) tracers
on the largest scales has recently received great interest as a probe
of inflation and its alternatives.  In the presence of primordial
non-Gaussianity, biased tracers can exhibit a significant scale-dependent
bias with respect to the matter distribution which increases strongly
towards large scales \cite{DalalEtal08}.  This can be used as a sensitive probe
of primordial non-Gaussianity \cite{SlosarEtal}.  Furthermore, ongoing, future, and proposed surveys
such as the Baryon Oscillation Spectroscopic Survey \cite{BOSS}, the Hobby Eberly Telescope Dark Energy Experiment \cite{hetdex}, HyperSuprime Cam, the Dark Energy Survey\footnote{\tt http://www.darkenergysurvey.org/}, the Subaru Prime Focus Spectrograph, BigBOSS \cite{BigBOSS},
the Large Synoptic Survey Telescope\footnote{\tt http://www.lsst.org/lsst},
the Wide-Field Infrared Survey Telescope\footnote{\tt http://wfirst.gsfc.nasa.gov/}, Euclid\footnote{\tt http://sci.esa.int/euclid/}, and the Square Kilometer Array\footnote{\tt http://www.skatelescope.org/} will 
probe modes that approach the comoving horizon.  All of this is strong
motivation to go beyond the Newtonian picture of galaxy clustering
widely adopted so far, and to embed this observable into a proper relativistic
context.  This is analogous to what has been done for the cosmic
microwave background (CMB), and several aspects have long been worked out
\cite{SW67,Sasaki87,BonvinEtal06}.  However, galaxy clustering involves
a few additional complications: first, it is intrinsically three-
rather than two-dimensional; second, one has to take into account
selection effects such as cuts on observed flux and redshift;
and third, the relation between intrinsic fluctuations in the galaxy
density and the fluctuations in the matter density is non-trivial.

In order to use the clustering of LSS tracers on scales approaching
the horizon $c/H(z)$, we thus need to understand the connection between
the theoretical predictions for cosmological perturbations and the
observationally inferred overdensities of galaxies.  In particular,
the perturbations in the metric, matter density, velocity, etc. 
are always defined
with respect to a particular choice of coordinates (gauge), whereas
observables should be independent of this gauge choice.  

Recently, \citet{yoo/etal:2009} have shown how to
calculate the observed galaxy overdensity in a generally covariant,
relativistic context.  In this paper, we perform a similar derivation,
generalizing their results in one important aspect.  On sufficiently large 
scales so that perturbations are linear (and assuming Gaussian initial 
conditions), one commonly assumes a linear relation between galaxy 
overdensities and perturbations in the matter density.  As we
show here however, the observed galaxy power spectrum depends on
which gauge these overdensities are referred to.  For example,
the linear bias assumed in \cite{yoo/etal:2009,yoo:2010} is
equivalent to a linear bias relation in
the constant-observed-redshift gauge;  on the other hand,
\cite{challinor/lewis:2011,BruniEtal} adopted a linear bias
in synchronous-comoving gauge.  

Clearly, this situation is not satisfactory, since the gauge choice
should not impact \emph{any observable} quantity.  However, we can
make progress using simple physical arguments.  In a universe with Gaussian 
adiabatic perturbations,
a galaxy knows about two properties of its large-scale environment:
the average, ``background'' density of matter, and the local age 
(growth history) of its environment.  
Thus, a general bias expansion should involve \emph{both} density and 
age (or local growth factor).  In this context,
the simplest gauge choice is the synchronous-comoving (sc) gauge, where
constant-time hypersurfaces are equivalent to constant-age
hypersurfaces.  Then, the bias with respect to age becomes irrelevant,
and we recover the well-known linear bias relation: 
$\d^{(\rm sc)}_g = b\:\d^{(\rm sc)}_m$.  Further advantages
of synchronous-comoving gauge are that the density field in N-body
as well as the output of commonly used Boltzmann codes
are given in this gauge.  We shall thus express
most of our results in synchronous-comoving gauge.  The transition
to other gauges can be performed easily using expressions given
in the appendices.  Note that when properly transformed, the results
derived in different gauges should agree.  

Recent papers by \citet{challinor/lewis:2011} and \citet{bonvin/durrer:2011}
provide a further reason to reinvestigate relativistic corrections to the
observed galaxy correlation, as they were not able to
reach agreement with the expressions given in Ref.~\cite{yoo/etal:2009}.  

The outline of the paper is as follows.  We begin by deriving
the observed galaxy density in terms of perturbations in the
synchronous-comoving gauge in \refsec{ng}.  In \refsec{bias}, we discuss
how galaxy biasing can be implemented in a gauge-invariant way.  
\refsec{Pkg} discusses the observed galaxy power spectrum.
We conclude in \refsec{concl}.  In the appendices, 
we present useful results on the conversion between different gauges
and metric conventions (\refapp{gTransf}),
more details on the derivations (\refapp{Pgderiv}), and
various analytical test cases for the expression
for the observed galaxy overdensity (\refapp{test}).  We also make
the connection with the work of
other recent papers \cite{yoo/etal:2009,yoo:2010,challinor/lewis:2011,bonvin/durrer:2011}
in \refapp{comp}.

%%%%%%%%%%%%%%%%%%%%%%%%%%%%%%%%%%%%%%%%%%%%%%%%%%%%%%%%%%%%%%%%%%%%%%%%%%%%%
%%%%%%%%%%%%%%%%%%%%%%%%%%%%%%%%%%%%%%%%%%%%%%%%%%%%%%%%%%%%%%%%%%%%%%%%%%%%%
%%%%%%%%%%%%%%%%%%%%%%%%%%%%%%%%%%%%%%%%%%%%%%%%%%%%%%%%%%%%%%%%%%%%%%%%%%%%%
\section{The observed galaxy density in synchronous gauge}
\label{sec:ng}

In this section, we compute the perturbations to the number density of 
LSS tracers in observed coordinates to linear order.

%%%%%%%%%%%%%%%%%%%%%%%%%%%%%%%%%%%%%%%%%%%%%%%%%%%%%%%%%%%%%%%%%%%%%%%%%%%%%
\subsection{Notation}

Throughout, unless otherwise noted, we adopt the synchronous-comoving
gauge and assume a flat background.  Specifically, we write
\be
ds^2 = a^2(\tau)\left\{ -d\tau^2 + \left[\, (1+2D)\,\d_{ij} + 2E_{ij} \right ]
dx^i\:dx^j
\right \},
\label{eq:g_sc}
\ee
where $\tau$ is the conformal time, $D$ is a scalar metric perturbation
while $E_{ij}$ is transverse and traceless, and 
related to the scalar perturbation $E$ by 
\be
E_{ij} = 
\left(\partial_i\partial_j - \frac{1}{3}\delta_{ij}\nabla^2 \right) E.
\ee
Latin letters denote spatial indices while Greek letters stand for
space-time indices.

In galaxy surveys, we observe the angular position of galaxies
as well as their redshift $\zt$.  Hereafter, we shall 
denote observed (or inferred) coordinates with a tilde.  
We can assign the galaxy
a position $\tilde\vx$ in three-dimensional cartesian coordinates
via
\be
\tilde\vx = \chit \hat{\tilde{\v{n}}},
\ee
where $\chit = \chib(\zt)$ and $\chib(z)$ is the distance-redshift relation 
in the background Universe\footnote{Throughout, we assume that we have
perfect knowledge of the background expansion history, and hence neglect
Alcock-Paczy\'nski-type distortions.  These can be taken into account
straightforwardly.}, and $\hat{\tilde{\v{n}}}$ is the unit vector 
in the direction of the observed position of the galaxy.  
In an unperturbed Universe, where both the observer and
the galaxy are comoving with the background matter, this position in fact 
corresponds to the true position (in comoving coordinates).  This 
is because the photon geodesic in an unperturbed Universe can be written
as
\be
\bar x^\mu(\chi) = (\tau_0 - \chi,\:\vnhatt\, \chi),
\label{eq:xmu}
\ee
where $\tau_0$ is the conformal time at observation, and we have chosen
the affine parameter to be the comoving distance (along the light ray).  
Thus, the space-time point of emission of the photon is given by
$\bar x^\mu(\chit)$.  

One convenient feature of this parametrization is that 
the geodesic equation with respect to $\chi$ coincides with that of 
conformally transformed coordinates with 
metric $\hat{g}_{\mu\nu}=g_{\mu\nu}/a^2$; in the case of a spatially flat FRW
universe, it is simply a straight line as in \refeq{xmu}.  The
corresponding affine parameter $\lambda$ in the physical FRW metric $g_{\mu\nu}$
is determined through $d\lambda/d\chi\propto a^2$.  

\begin{figure}[t!]
\centering
\includegraphics*[width=0.48\textwidth]{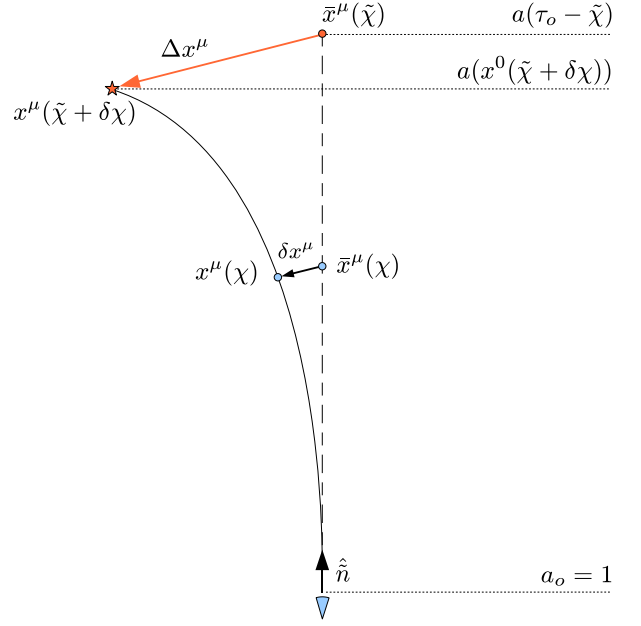}
\caption{Sketch of perturbed photon geodesics illustrating our notation. 
The observer is located at the bottom.  The solid line indicates the
actual photon geodesic tracing back to the source indicated by a star.  
The dashed line shows the apparent background photon geodesic tracing back to
an inferred source position indicated by a circle.}
\label{fig:sketch}
\end{figure}

In a perturbed Universe, the true location of the galaxy is defined by
the unique starting point of the geodesic which ends at the observer's location
[$\tilde\vx_o = (0,0,0)$], arrives out of the direction $\hat{\tilde{\v{n}}}$, 
and corresponds to a photon redshift $\zt$ (see \reffig{sketch}); 
in other words, the photon frequency at arrival at $\tilde\vx_o$ is
\be
\tilde\nu = \frac{\nu_0}{1+\zt}.
\ee
Here we assume we have some frequency standard $\nu_0$ (e.g., spectral
line) to compare the photon frequency to.  

In the following, it is useful to define projection operators,
so that for any spatial vector $X^i$ and tensor $E_{ij}$,
\ba
X_\parallel \equiv\:& \nhatt_i X^i, \vs
E_\parallel \equiv\:& \nhatt_i \nhatt_j E^{ij}, \vs
X^i_\perp \equiv\:& (\d^i_{\;j} - \nhatt^i \nhatt_j) X^j, {\rm~and}\vs
E_\perp \equiv\:& (\d^{ij} - \nhatt^i \nhatt^j) E_{ij}.\label{eq:proj1}
\ea
Note that for a traceless tensor $E_{ij}$, $E_\perp = -E_\parallel$.  
Correspondingly, we define projected derivative operators,
\ba
\partial_\parallel^i \equiv\:& \nhatt^i \nhatt^j \partial_j, \vs
\partial_\parallel \equiv\:& \nhatt^i\partial_i,{\rm ~and} \vs 
\partial_\perp^i \equiv\:& (\d^{ij} - \nhatt^i\nhatt^j)\partial_j
= \partial^i - \nhatt^i \partial_\parallel.%\vs
\label{eq:proj2}
\ea
Further, we find
\be
\partial_j \nhatt^i 
= \chit^{-1}(\delta_j^i - \nhatt^i\nhatt_j),
\ee
from which we derive a number of commutation relations. These include the commutators of the partial derivatives with $\nhatt$: 
\be
[\partial_i , \nhatt_j]
= \partial_i \nhatt_j
= \chit^{-1}(\delta_{ij} - \nhatt_i\nhatt_j) \label{eq:comm1}
\ee
and
\be
[\partial_{\parallel}, \nhatt_{i} ] =  [\nhatt^j \partial_j , \nhatt_i] 
= \nhatt^j [\partial_j , \nhatt_i] + [\nhatt^j,\nhatt_i] \partial_j
 = 0.
 \label{eq:comm-np}
\ee
We may also find the commutators of the derivative operators with each other. Those involving the parallel derivatives are
\begin{eqnarray}
[\partial_i , \partial_\parallel] &=& [\partial_i, \nhatt^j\partial_j]
\nonumber \\
&=& [\partial_i , \nhatt^j]\partial_j + [\partial_i,\partial_j]\nhatt^j
\nonumber \\
&=& \chit^{-1}(\delta_{ij} - \nhatt_i\nhatt_j)\partial_j + 0
\nonumber \\
&=& \chit^{-1}\partial_{\perp i}
\end{eqnarray}
and, since Eq.~(\ref{eq:comm-np}) shows that all components of $\nhatt$ commute with $\partial_\parallel$,
\be
[\partial_\parallel, \partial_{\parallel j}] = [\partial_{\parallel i}, \partial_{\parallel j}] = 0.
\ee
The perpendicular derivative satisfies
\be
[\partial_{\perp i} , \partial_{\parallel}]
= 
[\partial_{i}  - \partial_{\parallel i}, \partial_{\parallel}]
=
[\partial_i , \partial_{\parallel}]
= \chit^{-1}\partial_{\perp i}.
\label{eq:commn}
\ee
In these expressions, $\chit$ is the norm of the position vector so that $\nhatt^i = x^i/\chit$.
These relations are the analogue of the Christoffel symbols of spherical
polar coordinates.

We also define the projection of the Laplacian operator 
($\nabla^2 = \partial_i\partial^i$) as
\begin{eqnarray}
\partial^2_\parallel 
&\equiv&
\partial_{\parallel i}\partial_\parallel^i =
\partial_\parallel\partial_\parallel
{\rm ~and} \nonumber \\
\nabla^2_\perp
&\equiv& \partial_{\perp i}\partial_\perp^i
=\nabla^2 - \partial_\parallel^2 - \frac{2}{\chit}\partial_\parallel.
\end{eqnarray}
Finally, we make use of 
\be
\partial_i X^i = \partial_\parallel X_\parallel + \partial_{\perp\,i} X_\perp^i
+ X_\parallel \partial_i \nhatt^i \label{eq:divX}
\ee
and
\be
\partial_\parallel \nhatt^i = \nhatt^i\partial_{\perp i} = 0.
\ee
%\label{eq:dparnhat}

%!!!!!!!!!!!!!!!!!!!!!!!!!!!!!!!!!!!!!!!!!!!
%!!!!!!!!!!!!!!!!!!!!!!!!!!!!!!!!!!!!!!!!!!!
%%%%%%%%%%%%%%%%%%%%%%%%%%%%%%%%%%%%%%%%%%%%%%%%%%%%%%%%%%%%%%%%%%%%%%%%%%%%%
\subsection{Photon geodesics}

We can write the perturbed photon geodesic as
\be
x^\mu(\chi) = \bar x^\mu(\chi) + \d x^\mu(\chi); \label{eq:dxmu}
\ee
here $\d x^\mu$ is the perturbation to the photon path
(note that the value of the affine parameter at emission is no longer
given by $\chit$; see \reffig{sketch}).  
For our choice of affine parameter $\chi$, we
have in the unperturbed Universe [\refeq{xmu}]
\be
\frac{d\bar x^\mu}{d\chi} =  (-1, \vnhatt), 
\ee
while for the perturbed case [\refeq{dxmu}] we define
\be
\frac{d x^\mu}{d\chi} = (-1+\d\nu,\: \vnhatt + \d\v{e}).  
\ee
In terms of our affine parameter $\chi$, the first-order geodesic equations 
for the fractional frequency perturbation $\d\nu \equiv d\d x^0/d\chi$, 
and the fractional perturbations to the photon momentum 
$\d e^i \equiv d\d x^i/d\chi$ are
then given by
\be
\frac{d}{d\chi} \d\nu =  -(D' + E'_\parallel)\label{eq:geod_0}
\ee
and
\ba
\frac{d}{d\chi} \d e^i + 2\frac{d}{d\chi} \left (D \nhatt^i + E^i_{\;j} \nhatt^j\right) =\:& 
 D^{,i} + E_{jk}^{\;\;\;,i} \nhatt^j \nhatt^k \vs
 = D^{,i} + (E_\parallel)^{,i} 
-\:& \frac2\chi \left(E^i_{\; k}\nhatt^k - E_\parallel \nhatt^i\right).
\label{eq:geod_3} 
\ea 
Note that the projection onto $\nhatt$ and the spatial derivative do not
commute.  In the following, we denote $E_\parallel^{,i} \equiv (E_\parallel)^{,i}$,
i.e. in case of apparently ambiguous notation, the projection is taken before the derivative.
Here, primes denote derivatives with respect to $\tau$, and 
$d/d\chi = \partial_\parallel - \partial_\tau$.  
These equations are to be compared with Eqs. (9) and (12) in 
\cite{yoo/etal:2009}, respectively; note that 
$\chi_{\rm there} = -\chi_{\rm here}$ and $\d\nu_{\rm there}=-\d\nu_{\rm here}$.    

Before integrating Eqs.~(\ref{eq:geod_0},{\ref{eq:geod_3}), we need to determine the 
correct boundary conditions at the observer's position $\chi=0$.  
The observer's frame of reference is described by an orthonormal 
tetrad $e^a_\mu$.  In terms of this basis, the components of the 
unit vector $\vnhatt$ of the observed photon are given by
\be
\nhatt^a = \frac{e^a_\mu p^\mu}{p},
\label{eq:n1}
\ee
where $p^\mu$ is the observed photon momentum and $p = p_i p^i$.  
Using the metric \refeq{g_sc} together with the orthonormality condition
\be
g^{\mu\nu} e^a_\mu e^b_\nu = \eta^{ab},
\ee
we obtain
\ba
e^0_\mu =\:& (-1, 0, 0, 0) {\rm ~and}\vs
e^i_\mu =\:& \left(0,\: (1+D_o)\d^i_{\;j} + E^i_{o\,j}\right),
\label{eq:tetrad}
\ea
where a subscript $o$ indicates a quantity evaluated at the observer's 
position.
Inserting this into \refeq{n1}, and requiring that $\nhatt^a$ match the
observed direction of the photon, we obtain the following perturbations
to the photon momentum at the observer's position ($\chi=0$):
\be
\d\nu_o = 0 {\rm ~~and~~}
\d e_o^i = - D_o \nhatt^i - E^i_{o\,j} \nhatt^j.
\label{eq:deIC}
\ee
We now integrate the spatial component \refeq{geod_3},
enforcing \refeq{deIC} as boundary condition at $\chi=0$:
\ba
\d e^i(\chi) =\:& -2 (D \nhatt^i + E^i_{\;j}\nhatt^j)_\chi \vs
& + \int_0^\chi d\chi' \left[ D^{,i} + E_\parallel^{,i}
- \frac2{\chi'} \left(E^i_{\; k}\nhatt^k - E_\parallel \nhatt^i\right) \right]
\vs
& + (D\nhatt^i + E^i_{\;j} \nhatt^j)_o.
\ea
Integrating again up to $\chit$ yields the displacements $\d x^i$:
\ba
\d x^i(\chit) =\:& \int_0^{\chit} d\chi\;\d e^i \nonumber \\
 =\:& \chit( D \nhatt^i + E^i_{\;j} \nhatt^j)_o\vs
& + \int_0^{\chit}d\chi\: \Big\{\!\!   -2(D \nhatt^i + E^i_{\;j}\nhatt^j)
\nonumber \\
&\quad + (\chit-\chi) \Big[ D^{,i} + E_\parallel^{,i} 
 - \frac2{\chi}(E^i_{\; k}\nhatt^k - E_\parallel \nhatt^i)
\Big]
\Big \}.\nonumber
\ea
Integrating the time-component \refeq{geod_0} of the geodesic equation
yields
\ba
\d\nu(\chit) =\:& -\int_0^{\chit} d\chi\;(D' + E'_\parallel).  
\ea
This frequency shift contains the Doppler, Sachs-Wolfe, and
integrated Sachs-Wolfe effects \cite{SW67}, as shown in \refapp{ISW}.
Noting that $d x^0/d\chi = -1 + \d\nu$, we then obtain the time
delay
\be
\d x^0(\chit) = \int_0^{\chit} d\chi\; \d\nu
 = -\int_0^{\chit} d\chi\; (\chit-\chi)(D'+E'_\parallel).
\ee

In order to obtain the perturbations to the source position, we need
to relate the affine parameter at emission to the observed redshift.  
Recall that in synchronous-comoving gauge, the comoving observers' four-velocity
is given by $u_\mu = (a,0,0,0)$.  
Then, the redshift along the perturbed geodesic 
at affine parameter $\chi$ is given by
\ba
1 + z(\chi) =\:& \frac{(a^{-2}\,u_\mu\: dx^\mu/d\chi)_\chi}{(a^{-2}\,u_\mu\: dx^\mu/d\chi)_o}
\nonumber \\
=\:& \frac{[-1+\d\nu(\chi)]/a(x^0(\chi))}{-1}
= \frac{1+\d z}{a(x^0(\chi))} .
\label{eq:z1}
\ea
In the second line, we have set $a_o = 1$ and defined 
\be
\d z(\chit) \equiv -\d\nu(\chit) = \int_0^{\chit} d\chi\;(D'+E'_\parallel).
\label{eq:deltaz}
\ee
For a given source observed at redshift $\zt$, \refeq{z1} is an implicit
relation for the affine parameter $\chi_e$ at emission,
\be
1 + z(\chi_e) = 1 + \zt,
\ee
which defines the space-time location of the source through 
$x^\mu_{\rm source} = x^\mu(\chi_e)$.  Note that since the conformal 
time of emission is $\tau = x^0(\chi_e)$, the redshift $\bar z(\chi_e)$ that 
would have been observed for the same source without any perturbations 
along the line of sight is given by $1/[1+\bar z(\chi_e)] = a(x^0(\chi_e))$.  
Hence, \refeq{z1} at $\chi=\chi_e$ can also be written as
\be
1 + \zt = (1+\bar z) (1 + \d z(\chi_e)).  
\label{eq:dzdef}
\ee
To zeroth order (in the background), $z(\chi)=\bar z(\chi)$,
and hence $\chi_e = \chit$.  We can then expand
$\chi_e = \chit + \d\chi$, and \refeq{z1} at first order yields
\be
1 + \zt = (1 + \zt)[1 - (aH)_{\zt} (\d x^0 - \d\chi) + \d z].
\ee
Solving this for the perturbation to the affine parameter, we obtain
\be
\d\chi = \d x^0 - \frac{1+\zt}{H(\zt)} \d z.
\label{eq:dchi}
\ee

Finally, given \refeq{dxmu}, we can relate the observed position 
$\tilde\vx$, inferred assuming unperturbed geodesics $\bar x^\mu$, and the 
true position $\vx$ through (see \reffig{sketch})
\be
\vx = \tilde\vx + \Delta\vx = \tilde\vx + 
\chit \d\vnhatt + \frac{d\bar{\vx}}{d\chi}\d\chi.
\label{eq:dx}
\ee
Separating into longitudinal and perpendicular parts, we obtain
\be
\D x_\parallel = \d x^i(\chit)\nhatt_i + \d\chi =  \d x_\parallel + \d x^0
- \frac{1+\zt}{H} \d z \label{eq:Dx3}
\ee
and
\be
\D x^i_\perp =  \d x^i - \nhatt^i \d x_\parallel.
\label{eq:Dxperp}
\ee
\refeqs{Dx3}{Dxperp} can be further simplified to obtain
\be
\D x_\parallel = -\int_0^{\chit}d\chi\; (D+E_\parallel) - \frac{1+\zt}{H(\zt)}
\int_0^{\chit} d\chi\;(D' + E_\parallel')
\label{eq:Dx32}
\ee
and
\ba
\D x_\perp^i 
%=\:& \chit \left( E^i_{\;j}\nhatt^j - E_\parallel \nhatt^i\right)_o
%\vs
%& + \int_0^{\chit} d\chi 
%\Big[\! -2\left( E^i_{\;j}\nhatt^j - E_\parallel \nhatt^i\right) \vs
%& \hspace*{1.6cm}
%+ (\chit-\chi)\partial_\perp^i\left(D + E_\parallel\right) \vs
%& \hspace*{1.6cm}
%-2 \frac{\chit-\chi}{\chi} \left(E^i_{\; k}\nhatt^k - E_\parallel \nhatt^i\right)
%\Big]\vs
=\:& \chit \left( E^i_{\;j}\nhatt^j - E_\parallel \nhatt^i\right)_o \vs
& + \int_0^{\chit} d\chi 
\Big[\! -2 \frac{\chit}{\chi} \left( E^i_{\;j}\nhatt^j - E_\parallel \nhatt^i\right) \vs
& \hspace*{1.6cm}
+ (\chit-\chi)\partial_\perp^i\left(D + E_\parallel\right) 
\Big].
\label{eq:Dxperp2}
\ea
Note that the terms involving perturbations at the observer's location
have dropped out of \refeq{Dx32}.  This equation does not quite agree 
with Eq.~(16) in \cite{yoo/etal:2009}, where $E_\parallel$ has 
the opposite sign in the first term.  This difference also carries 
through to their Eq.~(36).  
All these terms come from the metric perturbation 
$\d g_{ij}\nhatt^i\nhatt^j$ however, hence they should
always involve the combination $D+E_\parallel$.  For the numerical results 
reported in \cite{yoo/etal:2009,yoo:2010}, this difference is of no
relevance as they evaluate the 
power spectrum in conformal Newtonian gauge where $E=0$.

%%%%%%%%%%%%%%%%%%%%%%%%%%%%%%%%%%%%%%%%%%%%%%%%%%%%%%%%%%%%%%%%%%%%%%%%%%%%%
\subsection{Observed galaxy number density}

The observed number of galaxies contained within a volume $\tilde V$
defined in terms of the observed coordinates is given by a
(gauge-invariant) integral over a three-form
\ba
N =\:& \int_{\tilde V} \sqrt{-g(x^\alpha)}\: n_g(x^\alpha)\: \varepsilon_{\mu\nu\rho\sigma}
u^\mu(x^\alpha)
\frac{\partial x^\nu}{\partial \tilde x^1}
\frac{\partial x^\rho}{\partial \tilde x^2}
\frac{\partial x^\sigma}{\partial \tilde x^3}
d^3\tilde\vx \nonumber\\
 =\:& \int_{\tilde V} \sqrt{-g}\: n_g(x^\alpha)\:\frac{1}{a(x^0)} \varepsilon_{ijk}
\frac{\partial x^i}{\partial \tilde x^1}
\frac{\partial x^j}{\partial \tilde x^2}
\frac{\partial x^k}{\partial \tilde x^3}
d^3\tilde\vx \nonumber\\
=\:& \int_{\tilde V} \sqrt{-g}\: n_g(x^\alpha)\:\frac{1}{a(x^0)}\: 
\left |\frac{\partial x^i}{\partial \tilde x^j}\right|
d^3\tilde\vx.\label{eq:Ntilde}
\ea
Here, $\vx$ is given in terms of $\tilde\vx$ by \refeq{dx}, 
$\varepsilon_{\mu\nu\rho\sigma}$ is the Levi-Civita tensor, and 
$n_g$ is the physical number density of galaxies as a function of
``true'' comoving locations (in synchronous-comoving gauge).  
In the second line, we have adopted the synchronous-comoving gauge, 
where the oberver velocities reduce to $u^\mu = (1/a,0,0,0)$.  
In this case, not surprisingly, the volume element reduces to the purely
spatial Jacobian 
$|\partial x^i/\partial\tilde x^j|$.  Note that perturbations enter
\refeq{Ntilde} in three places: through the determinant $\sqrt{-g}$; 
through the position- and redshift-dependence of the galaxy density $n_g$,
and through the Jacobian $|\partial x^i/\partial\tilde x^j|$.  
This Jacobian is
\be
\left|\frac{\partial x^i}{\partial\tilde x^j}\right| = 
\left| \d^i_j + \frac{\partial \D x^i}{\partial\tilde x^j} \right|
= 1 + \frac{\partial \D x^i}{\partial\tilde x^i},
\ee
where we have worked to first order in the displacements $\D\vx$.
Furthermore, noting that $\sqrt{-\bar g} = a^4$, where $\bar g_{\mu\nu}$ is the
background metric, we have
\be
\sqrt{-g} = a^4\left(1 + \frac{1}{2} \d g^\mu_{\;\mu}\right).
\ee
Finally, the galaxy density perturbations are usually measured with respect
to the average density of galaxies at fixed observed redshift, 
$\bar n_g(\zt)$.  We assume that when averaged over the whole survey,
$\<\d z\> = 0$ so that $\<\zt\> = \bar z$ [\refeq{dzdef}].  
Also, in this paper, we follow common convention and
define the galaxy density perturbations $\d_g$ with respect to the
\emph{comoving} galaxy density.  We thus have for the 
intrinsic comoving galaxy density
\be
a^3(\bar z) n_g(\vx, \bar z) = a^3(\bar z) \bar n_g(\bar z)\:[1 + \d_g(\vx, \bar z)],
\ee
where $\bar z$ again denotes the redshift that would have been observed in an
unperturbed Universe, and $\d_g$ denotes the intrinsic fluctuations 
in the comoving galaxy density.  Using \refeq{dzdef} and 
expanding to first order, we obtain
\ba
a^3(z) n_g(\vx, \bar z) =\:& a^3(\zt) \bar n_g(\zt)\:[1+\d_g(\tilde\vx)] 
\nonumber \\
& - (1+\zt) \frac{d (a^3 \bar n_g)}{dz}\Big|_{z=\zt} \d z.
\label{eq:npp2}
\ea
Note that the distinction between 
$\d_g(\tilde\vx)$ and $\d_g(\vx)$ is second order (this effect, analogous
to CMB lensing, can however become important for rapidly varying
correlation functions \cite{DodelsonEtal08}).  

We can now expand \refeq{Ntilde}.  We define the \emph{observed}
galaxy density $\tilde n_g$ via 
\be
\int_{\tilde V} a^3(\zt) \tilde n_g(\tilde\vx,\zt) d^3\tilde\vx = N,
\ee
so that
\ba
a^3(\zt)\tilde n_g(\tilde \vx,\zt) =\:&  \sqrt{-g} \frac{1}{a(\bar z)} 
\: n_g(\vx, \bar z)\:
\left |\frac{\partial x^i}{\partial \tilde x^j}\right|\label{eq:nptilde}\\
=\:& \left(\!1 + \frac{1}{2}\d g^\mu_{\;\mu}\!\right) a^3(\bar z)
n_g(\vx,\bar z) \left (1 + \frac{\partial\D x^i}{\partial\tilde x^i}\right)\!.
\nonumber
\ea
For the Jacobian, we use \refeq{divX} to obtain
\ba
\frac{\partial\D x^i}{\partial \tilde x^i}
=\:& \partial_\parallel \D x_\parallel + \D x_\parallel \partial_i\nhatt^i 
+ \partial_{\perp\,i} \D x^i_\perp \vs
=\:& \partial_\parallel \D x_\parallel + \frac{2\D x_\parallel}{\chit}
 - 2 \hat\k, \label{eq:J}
\ea
where we have defined the coordinate convergence as 
\be
\hat\k \equiv -\frac{1}{2} \partial_{\perp\,i} \D x_\perp^i.
\label{eq:khat}
\ee
Note that the coordinate along $\vnhatt$ is defined through the observed 
redshift $\zt$. Hence, 
$\partial_\parallel = (d\bar\chi/dz) \partial/\partial\zt$.  Since the
derivative is applied to first-order displacements, it suffices to use the
zeroth order expression 
$\partial_\parallel = \partial/\partial\chi|_{\chi=\chit}$.  

Using \refeq{Dx32} for $\D x_\parallel$, we obtain the first two terms
in the Jacobian
\be
\D x_\parallel = -\int_0^{\chit} d\chi (D + E_\parallel)
-\frac{1+\zt}{H(\zt)} \d z
\label{eq:dxpara}
\ee
and
\ba
\partial_\parallel \D x_\parallel =\:& -(D+E_\parallel)\Big|_{\chit}
- H(\zt) \d z \frac{d}{d\zt}
\left[\frac{1+\zt}{H(\zt)}\right] \vs
& - \frac{1+\zt}{H(\zt)} (D' + E'_\parallel).
\ea
Using \refeq{Dxperp2} we find for the convergence
\be
\hat\k = \k + \frac{1}{2} \nabla_{\perp}^2 E
 + \frac1{\chit}E_{,i}(o)\hat n^i 
-\hat n^i E'_{,i}(o),
\label{eq:khat2}
\ee
where 
\be
\k = 
- \frac{1}{2}
\int_0^{\chit}
d\chi
\left(\chit-\chi\right)
\frac{\chi}{\chit}
\nabla^2_\perp 
\left(D - \frac{1}{3}\nabla^2 E + E''\right)
\label{eq:kappadef}
\ee
is the usual definition of the convergence in synchronous-comoving gauge
and the derivatives of $E$ with $(o)$ are evaluated at the observer.
The details of the derivation of $\hat\k$ can be found in \refapp{kappa}.

Finally, we can expand \refeq{nptilde} to linear order in the perturbations:
\ba
\frac{\tilde n_g(\tilde\vx,\zt)}{\bar n_g(\zt)} =\:& 
1 + 3D + \d_g 
+ \bt \d z
 + \frac{2\D x_\parallel}{\chit} + \partial_\parallel\D x_\parallel
- 2\hat\k \vs
 =\:& 1 + 2 D - E_\parallel + \d_g 
+ \bt \d z
 + \frac{2\D x_\parallel}{\chit} - 2\hat\k \vs
&\!
- \left[1 - \frac{1+\zt}{H}\frac{d H(\zt)}{d\zt}
\right] \d z
 - \frac{1+\zt}{H(\zt)} (D' + E'_\parallel)\Big|_{\chit}
.\label{eq:nobssc}
\ea
Here, $\d\chi$ is given by \refeq{dchi}, and $\d z$ is given by 
\refeq{deltaz}.  Further, we have defined
\be
\bt \equiv \frac{d\ln (a^3\bar n_g)}{d\ln a}\Big|_{\!\zt} 
= - (1+\zt)\frac{d\ln (a^3\bar n_g)}{dz}\Big|_{\!\zt}.
\label{eq:btaudef}
\ee
Throughout this section, we have neglected the magnification bias
contribution.  This will be discussed in \refsec{mag}.

%%%%%%%%%%%%%%%%%%%%%%%%%%%%%%%%%%%%%%%%%%%%%%%%%%%%%%%%%%%%%%%%%%%%%%%%%%%%%
\subsection{Observed galaxy density contrast}

In galaxy surveys, we calculate the galaxy density perturbations
$\tilde\delta_g(\tilde\vx) = \tilde n_g(\tilde\vx)/\left<\tilde n_g\right>-1 $ 
by referring to the
average number density $\<\tilde n_g\>$ at \emph{fixed observed redshift}.  
In this sense, we measure the galaxy density contrast 
in the uniform redshift gauge, where the constant-time hypersurface is 
defined by $\delta z = 0$.  
In the following, we assume that all fluctuations in $n_g$ are
due to large-scale structure, and ignore any contributions from
e.g. varying survey depth and extinction.  Using the result \refeq{nobssc},
we obtain
\ba
\nonumber
\tilde{\delta}_g(\tilde\vx)
=\:& \d_g + \bt\d z
- \frac{1+\zt}{H(\zt)} \partial_\parallel^2E'
\vs
&- \left(
1 - \frac{1+\zt}{H}\frac{d H(\zt)}{d\zt}
+
\frac{2}{\chit}\frac{1+\tilde{z}}{H(\tilde{z})}
\right) \d z 
+2\phi 
\vs
&
-\frac{2}{\chit}[E'-E'(o)]
-\frac{2}{\chit}
\int_0^{\tilde\chi}d\chi
\left(\phi + E''\right) - 2\k
\vs &
-\frac{1+\tilde{z}}{H(\tilde{z})}\phi'
+ 2\partial_\parallel E'(o),
\label{eq:dgsc}
\ea
where $\phi \equiv D - \nabla^2 E/3$ (see \refapp{derivdg} for more
details). In \refapp{test}, we apply \refeq{dgsc} to analytical test
cases where the exact result is known;  these results serve as a cross-check
of our result as well as to elucidate the significance of the various contributions.
The terms in the first line contain the gauge-invariant intrinsic
galaxy density perturbation (\refsec{bias}) and the standard
redshift-distortion contribution.  The terms on the second line
contain the change in volume entailed by the redshift perturbation.  
Finally, the last two lines contain further volume distortions from
the metric at the source position, Doppler effect, time delay,
and lensing convergence.

The observer terms in Eq.~(\ref{eq:dgsc}), $2\tilde\chi^{-1}E'(o)$ and $2\partial_\parallel E'(o)$, contribute only to the monopole and dipole of the galaxy distribution respectively. Therefore for most analyses that use the $\ell\ge 2$ multipoles of the galaxy distribution, or the small angle approximation, they can be neglected. The monopole term is not even measurable since we do not know the true mean galaxy density. (The dipole of a galaxy distribution {\em is} measurable in principle; indeed it has been used to search for e.g. inhomogeneous initial conditions \cite{hirataquasar}.)

In \refapp{comp} we connect this expression with the result
for $\d_{\rm obs}$ of \cite{yoo/etal:2009}.  
Besides notational differences we clarify there, 
the most critical difference is that we obtain a term 
\be
\d_g + \bt \delta z,
\label{eq:newterm}
\ee
which takes into account the difference in the mean galaxy number 
density [\refeq{npp2}] between synchronous-comoving gauge and 
the uniform-redshift slicing, on which we measure the 
mean number density [the combination \refeq{newterm} is
manifestly gauge invariant, as we shall show in the next section].   
This term is not considered in \citet{yoo/etal:2009},
as they relate the galaxy overdensity
to the matter overdensity $\d_m$ through the gauge-invariant relation
\begin{equation}
\delta_g = b(\delta_m-3\delta z),
\label{eq:dgY}
\end{equation}
where $\d_m(\vx) = \rho_m(\vx)/\bar\rho_m-1$ is the fractional matter
overdensity defined in whichever gauge is adopted.  Clearly, \refeq{dgY}
is equivalent to assuming a linear bias relation in terms of density,
$\delta_g=b\delta_m$, in the \emph{uniform-redshift gauge}.  
We will discuss these issues in the next section.  

On the other hand, by transforming \refeq{dgsc} into conformal-Newtonian gauge,
we are able to confirm that our result matches Eq.~(30) in the recent paper
by Challinor \& Lewis \cite{challinor/lewis:2011}.  This comparison
is detailed in \refapp{comp} as well.

\subsection{Magnification bias}
\label{sec:mag}

Equation~(\ref{eq:dgsc}) applies to the clustering of objects selected according to their intrinsic physical properties, their redshift, and their observed position on the sky. While one could in principle construct such a sample -- e.g. a temperature-limited sample of X-ray clusters -- most real samples in observational cosmology also depend on the apparent flux from the source. That is, they have a selection probability that depends on how the luminosity distance $D_{\rm L}$ differs from the mean luminosity distance $\bar D_{\rm L}(z)$. The sample selection may also depend on the angular size of the source, but this is not independent since the conservation of photon phase space density relates the angular diameter and luminosity distances\footnote{In the presence of opacity of intergalactic medium (IGM), e.g. due to Thomson scattering, this is not necessarily true. However, IGM opacity is a very small effect and we do not consider it in this paper.}
\begin{equation}
D_{\rm L} = (1+\tilde z)^2D_{\rm A}.
\end{equation}
This section evaluates the additional terms that appear in Eq.~(\ref{eq:dgsc}) in the presence of a dependence on magnification.  
%Such terms are however highly dependent on the galaxy sample and are not included in the numerical or graphical results displayed here. Their implications for realistic surveys on very large scales (and in particular how they differ from magnification effects in the small-scale limit) are left for a future paper.
The key parameter that we need to measure is the magnification
\be
\mmu \equiv \frac{D^{-2}_{\rm A}}{\bar D_{\rm A}^{-2}(\tilde z)} = \frac{D^{-2}_{\rm L}}{\bar D_{\rm L}^{-2}(\tilde z)},
\ee
which has mean value 1 and represents the perturbation to the solid angle or flux of a source, relative to a source at the same observed redshift $\tilde z$ in the unperturbed Universe. We may also write $\delta\mmu\equiv\mmu-1$. The galaxy overdensity is then
\be
\tilde\delta_g = \tilde\delta_g({\rm no~mag}) + \Q\delta\mmu,
\label{eq:dg-corr}
\ee
where $\tilde\delta_g$(no~mag) is the overdensity computed from Eq.~(\ref{eq:dgsc}) and
\be
\Q = \left.\frac{\partial\ln \tilde n_g}{\partial\ln\mmu}\right|_{\tilde z}
\ee
is the dependence of the observed number counts on magnification. For a magnitude-limited sample with cumulative luminosity function $\bar n(>L)$ we have $\Q=-d\ln\bar n(>L)/d\ln L$. A more general criterion, e.g. one that includes a ``size'' cut to reject stellar contamination
\cite{schmidtetal09}, would have $\Q$ that must be determined by simulating the observations.

From Eq.~(\ref{eq:dg-corr}) we see that we need only determine how $\delta\mmu$ depends on the metric perturbations in order to have a complete description of the magnification bias.

Fortunately, we have already constructed the key ingredients in evaluating $\mmu$. If we consider a right-handed 3-dimensional orthonormal basis $\{\nhatt^i, \hat \alpha^i, \hat \beta^i\}$ where $\nhatt^i$ is the direction of observation, then the angular diameter distance can be inferred from the area perpendicular to the line of sight spanned by the rays along the past light cone near $\nhatt^i$,
\be
D_A^2 = \sqrt{-g(x^\alpha)} \,\varepsilon_{\mu\nu\rho\sigma} u^\mu\hat\ell^\nu \frac{\partial x^\rho}{\partial \nhatt^i} \frac{\partial x^\sigma}{\partial\nhatt^j} \hat\alpha^i \hat\beta^j,
\ee
where the partial derivatives are taken at fixed $\tilde\chi$ and $\hat\ell^\mu$ is a unit purely spatial vector ($u_\mu\hat\ell^\mu=0$) pointed away from the observer (in the sense that a photon emitted from the source with 4-velocity parallel to the null direction $u^\mu-\ell^\mu$ reaches the observer). Using the chain rule to replace the derivatives with those involving $\tilde x^i$ gives
\begin{eqnarray}
\mmu^{-1} &=& \frac{D_A^2}{[\bar a(\tilde\chi)]^2\tilde\chi^2}
\nonumber \\
&=& \frac{\sqrt{-g(x^\alpha)}}{[\bar a(\tilde\chi)]^2} \,\varepsilon_{\mu\nu\rho\sigma} u^\mu\hat\ell^\nu \frac{\partial x^\rho}{\partial \tilde x^i} \frac{\partial x^\sigma}{\partial\tilde x^j} \hat\alpha^i \hat\beta^j.
\end{eqnarray}
Next we observe that $\hat\ell^\nu$ is parallel to the spatial part of $L^\nu = \partial x^\nu/\partial\tilde\chi |_{\hat{\bf n}}$, since it is the spatial part of the tangent vector to the past light cone. Since $\hat\ell^\nu$ is a unit vector and $L^\nu$ is a past-directed null vector, it follows that
\be
\hat\ell^\nu = \frac{L^\nu}{L^\sigma u_\sigma} + u^\nu;
\ee
using $u_\sigma=(-a,0,0,0)$ and $u^k=0$ we find
\be
\hat\ell^k = \frac{L^k}{a(x^0)\,L^0} = \frac1{a(x^0)} \frac{\partial x^k/\partial\tilde\chi}{-\partial x^0/\partial\tilde\chi}.
\ee
Using this and collapsing the Levi-Cevita symbol to 3 dimensions, we find
\be
\mmu^{-1} = \frac{\sqrt{-g(x^\alpha)} }{-[\bar a(\tilde\chi) a(x^0)]^2\partial x^0/\partial\tilde\chi}
\varepsilon_{abc}\frac{\partial x^a}{\partial\tilde x^h}
 \frac{\partial x^b}{\partial \tilde x^i} \frac{\partial x^c}{\partial\tilde x^j} \nhatt^h \hat\alpha^i \hat\beta^j.
\ee
Since $\{\nhatt^i, \hat \alpha^i, \hat \beta^i\}$ form an orthonormal basis, we simplify this to
\be
\mmu^{-1} = \frac{\sqrt{-g(x^\alpha)} }{-[\bar a(\tilde\chi) a(x^0)]^2 \partial x^0/\partial\tilde\chi} \left| \frac{\partial x^i}{\partial\tilde x^j} \right|.
\label{eq:Minv}
\ee
Now we are in a position to compute the pieces of Eq.~(\ref{eq:Minv}). We already know that $\sqrt{-g(x^\alpha)} = a^4(1+3D)$ and $|\partial x^i/\partial\tilde x^j| = 1+\partial\D x^i/\partial\tilde x^i$ [the latter is given by Eq.~(\ref{eq:J})]. Finally the null condition gives
\be
-\frac{dx^0}{d\tilde\chi} = (1+D+E_\parallel)\frac{dx_\parallel}{d\tilde\chi} = 1+D+E_\parallel+\partial_\parallel \Delta x_\parallel.
\ee
We thus find
\be
\mmu^{-1} = \left[\frac{a(x^0)}{\bar a(\tilde\chi)}\right]^2 \left(1+2D - E_\parallel + 2\frac{\Delta x_\parallel}{\tilde\chi} - 2\hat\k\right),
\ee
or
\be
\delta\mmu = -2 \delta z -2D + E_\parallel - 2\frac{\Delta x_\parallel}{\tilde\chi} + 2\hat\k.
\ee
This makes sense: the perturbation to the magnification contains the obvious coordinate convergence term $2\hat\k$, but it also has three other pieces: a contribution $-2\Delta x_\parallel/\tilde\chi$ associated with bringing the source closer to or farther from the observer; a contribution $2(-\delta z-D)$ associated with the isotropic conversion from coordinate distances to physical distances (itself having both a part from the change in scale factor at the source and the metric perturbation); and a part $E_\parallel$ associated with the anisotropy of the coordinate system ($E_{ij}$ is traceless and magnification depends only on the perturbation to transverse distances).

Expanding $\Delta x_\parallel$ using Eq.~(\ref{eq:dxpara}) and $\hat\k$ using Eq.~(\ref{eq:khat2}) gives an alternate expression
\begin{eqnarray}
\delta\mmu &=& -2\delta z-2\phi -\frac2{\tilde\chi}[\partial_\parallel E-\partial_\parallel E(o)] + 2\k - 2\partial_\parallel E'(o)
\nonumber \\ &&
+ \frac2{\tilde\chi}\left[ \int_0^{\tilde\chi}(D+E_\parallel)d\chi + \frac{1+\zt}{H(\zt)}\delta z \right].
\end{eqnarray}
The integral may be simplified by replacing $D+E_\parallel \rightarrow \phi + \partial_\parallel^2E$ and then doing a double integration by parts using Eq.~(\ref{eq:integ-dip}),
\begin{eqnarray}
\delta\mmu &=& -2\phi + \frac2{\tilde\chi} [E' -E'(o)] + 2\k - 2\partial_\parallel E'(o)
\nonumber \\ &&
+ \frac2{\tilde\chi} \int_0^{\tilde\chi}(\phi + E'')d\chi + \left[-2+\frac2{\tilde\chi}\frac{1+\zt}{H(\zt)}\right]\delta z.~~~
\label{eq:magsc}
\end{eqnarray}

%%%%%%%%%%%%%%%%%%%%%%%%%%%%%%%%%%%%%%%%%%%%%%%%%%%%%%%%%%%%%%%%%%%%%%%%%%%%%
%%%%%%%%%%%%%%%%%%%%%%%%%%%%%%%%%%%%%%%%%%%%%%%%%%%%%%%%%%%%%%%%%%%%%%%%%%%%%
%%%%%%%%%%%%%%%%%%%%%%%%%%%%%%%%%%%%%%%%%%%%%%%%%%%%%%%%%%%%%%%%%%%%%%%%%%%%%
\section{Galaxy bias in a relativistic context}
\label{sec:bias}

In order to make progress from \refeq{dgsc}, we need to relate the
intrinsic galaxy overdensity $\d_g$ (here written in synchronous-comoving gauge)
to the matter and metric perturbations.  Fortunately, since we are interested
in large scales, we only need to consider terms linear in perturbations.  
What are the relevant quantities on which the physical galaxy density
might depend? 
 
The most important characteristics of the large-scale 
environment of a given galaxy are its mean density, and the evolutionary
stage (proper time since the Big Bang, or linear growth factor).  
In fact, on sufficiently large scales, these are the only quantities
of relevance to the galaxy two-point correlations \cite{McDonaldRoy}
\footnote{This assumes that there is no orientation-dependent selection
of galaxies.  Such a selection will introduce a dependence on the
large-scale tidal field as well \cite{Hirata09}.}.  

We can formalize this statement by considering some large spatial volume
within the Universe centered around the spacetime point $x_p^\mu$, on 
a constant-age hypersurface, $t_U =\:$constant,
where $t_U$ denotes the proper time of comoving observers since the
Big Bang.  
Then, the number of galaxies 
(or, more generally, tracers) within that
volume can only depend on the enclosed mass $M$, and the age of the
Universe in that volume $t_U$ which is being kept fixed:
\be
N_g = F(M;\: t_U;\: x_p^\mu).
\label{eq:FofM}
\ee
Here, the explicit dependence on $x_p^\mu$ indicates any stochasticity
in the relation between $N_g$ and the local density and age.  
We now assume that the volume $V$ is large enough so that linear perturbation
theory applies.  Then, in a given coordinate system $(\tau,\vx)$, the 
enclosed mass is given by
\ba
M =\:& \int_V \rho = \rhob(\tau) [1 + \d\ln\rho] V\vs
=\:& \rhob(\tau) [ 1 + \d_m - 3 a H \d\tau ] V.\label{eq:M}
\ea
Here, $\rhob(\tau)$ is the average (physical) matter density in the background
(equivalent to $\rho$ averaged over the entire constant-coordinate-time 
hypersurface), while $\d_m$ is the matter density perturbation on 
a \emph{constant-coordinate-time} hypersurface.  
The second line follows from $\d\ln\rho \equiv  \rho/\rhob - 1$ and
$d\ln\rhob/d\tau = -3aH$, and we have defined $-\d\tau(\vx)$ to be 
the displacement in coordinate time corresponding to a $t_U=\:$constant 
hypersurface:
\be
a(\tau) [\tau - \d\tau(\vx)] = t_U = \rm constant.
\ee
Thus, the term $-3aH\d\tau$ in \refeq{M} comes in from going from a 
constant-age hypersurface to a constant-coordinate-time hypersurface.  
Note that in \refeq{M} the perturbations are to be considered averaged 
over the volume $V$.  

In exactly the same way, we can define the average (physical) galaxy number
density $\bar n_g$ on constant-coordinate-time hypersurfaces.  
The same reasoning leading to \refeq{M} yields
\ba
N_g =\:& \int_V n_g = \bar n_g(\tau) [1 + \d\ln n_g] V\vs
=\:& \bar n_g(\tau) \left[ 1 + \d_g + \btp\: a H\d\tau \right] V,\label{eq:Ng}
\ea
where $\btp = d\ln\bar n_g/d\ln a$.  
We can now equate this to our general ansatz \refeq{FofM},
\ba
N_g =\:& F(M;\:t_U;\:x_p^\mu) \vs
=\:& \bar F(\rhob V; t_U) [ 1 + b (\d_m - 3 a H \d\tau) + \eps] \vs
=\:& \bar n_g(\tau) \left[ 1 + \d_g + \btp\:a H\d\tau \right] V.\label{eq:Ng2}
\ea
Here, we have defined $\bar F(M, t_U) \equiv \< F(M, t_U, x_p^\mu) \>_{t_U}$,
and introduced the bias
\be
b \equiv \frac{\partial\ln \bar F(M; t_U)}{\partial\ln M}\Big|_{\rhob V}
=\: \frac{\partial\ln \bar F(\rho_V V; t_U)}{\partial \ln\rho_V}\Big|_{\rhob}
\label{eq:bdef}
\ee
and the stochastic contribution to galaxy density
\be
\eps(x^\mu) = \frac{F(M, t_U, x^\mu)}{\bar F(M, t_U)} - 1.
\label{eq:epsdef}
\ee
Also $\rho_V$ denotes the average matter density (on the $t_U=$~const slice) 
within the volume $V$.

At first order, the bias $b$ defined in this way is only a function of $\tau$. 
The stochastic contribution $\eps$ to galaxy clustering
is only a function of the spacetime point 
($\eps$ is here considered to be first order as well).  
In the background ($\d_m\to 0, \d\tau \to 0$), \refeq{Ng2} implies, not 
surprisingly, $\bar n_g(\tau) V = \bar F(\rhob V;a\tau)$.  To first order
in the perturbations, recall that \refeq{Ng2} must hold in \emph{any} 
coordinate system.  Thus, we conclude that the galaxy density
perturbation is given in general by
\ba
\d_g(x^\mu) =\:& b(\tau) [\d_m(x^\mu) - 3 a H(\tau) \d\tau(x^\mu)] \vs
& - \btp(\tau)\:aH(\tau) \d\tau(x^\mu)
+ \eps(x^\mu).
\label{eq:bias1}
\ea
On sub-horizon scales, $a H\d\tau$ becomes negligible compared to $\d_m$
(for standard choices of gauge).  
\refeq{bias1} shows that in this limit we recover the usual bias relation
$\d_g = b \d_m + \eps$.  Furthermore, if we choose synchronous gauge where
all comoving observers are synchronized so that $t_U = a \tau$ everywhere
and thus $\d\tau=0$, the linear bias relation holds on \emph{all} scales.  
Note that the definition \refeq{bdef} is precisely what is commonly
called a peak-background split bias parameter \cite{Kaiser84,MoWhite}.    

This derivation was phrased in terms of the physical galaxy density.  
The reasoning and \refeq{bias1} trivially hold for the comoving 
galaxy density $a^3 n_g$ as well, the only difference being
that $\d_g$ is now the fractional perturbation in comoving number density,
and $\btp$ is replaced with \refeq{btaudef},
\be
\bt = \frac{d\ln (a^3\bar n_g)}{d\ln a}.  \nonumber
\ee
From now on, we shall exclusively consider comoving number densities,
as we did in \refsec{ng}.    

The bias relation \refeq{bias1} holds in all gauges, and the bias parameters
$b$ and $\bt$ do not depend on the gauge choice.  
This is not very surprising, since these parameters are in principle
observable: $b$ [\refeq{bdef}] quantifies the response of 
the galaxy number in a given volume at fixed age of the Universe 
to a change in the average 
mass density (or enclosed mass) within this volume; $\bt$ [\refeq{btaudef}] 
quantifies the dependence of the average (background) number density of 
galaxies on the age of the 
Universe.  To see the gauge-invariance of these bias parameters 
explicitly, consider the effect 
of a change in the time coordinate,
\be
\tau \rightarrow \check\tau = \tau + T,
\ee
where $T$ can in general be a function of $\tau$ and $\vx$
(spatial gauge transformations do not affect the density perturbations at 
linear order).  
By using \refeq{gTscalar} in \refapp{gTransf}, we find that $\d_m$ and 
$\d_g$ transform as
\ba
\check\d_m =\:& \d_m + 3 a H T
{\rm~~and}\nonumber \\
\check\d_g =\:& \d_g - \frac{d\ln (a^3 \bar n_g)}{d\tau} T 
= \d_g - \bt\:a H T.
\ea
On the other hand, $\eps$ is gauge-invariant.  
Note that a change in time coordinate (slicing) implies a change
in the redshift perturbation $\d z$ [\refeq{dzdef}] through 
\be
(1+\check z)\d \check z = (1+z)\d z + H T,
\ee
with $d\bar z/d\tau=-H$.  

The previous two equations clearly show that the combination
$\d_g + \bt \d z $ appearing in \refeq{dgsc} is 
gauge-invariant, independent of any bias relation.  On the other 
hand, under the same gauge transformation with fixed $b$ and $b_e$, 
\refeq{bias1} changes as
\ba
\check\d_g =\:& b (\d_m - 3 a H \d\tau) - \bt\:a H (\d\tau + T) 
+ \eps \vs
 =\:& \d_g - \bt\: a H T,
\ea
where we have used \refeq{btaudef} in the second line.  
That is, we recover the gauge transformation of $\d_g$ with fixed 
bias parameters, and, in this sense,
the bias parameters defined through \refeq{bdef} and \refeq{btaudef} are
gauge-invariant.  

We now see that a gauge-invariant expression for the 
galaxy number density [\refeq{dgsc}], and a gauge-invariant definition of 
the galaxy bias are separate issues.  
In Refs.~\cite{yoo/etal:2009,yoo:2010,bonvin/durrer:2011},
the second issue was not addressed explicitly, and in
Refs.~\cite{yoo/etal:2009,yoo:2010} $\bt$ was implicitly set to zero.

%%%%%%%%%%%%%%%%%%%%%%%%%%%%%%%%%%%%%%%%%%%%%%%%%%%%%%%%%%%%%%%%%%%%%%%%
\subsection{Bias parameters from universal mass function approach}
\label{sec:univ}

In this section, we show how both $b$ and $\bt$ can be estimated
in the universal mass function approach.  We adopt the synchronous-comoving
gauge, which is implicit in the reasoning of this approach. The universal mass function
approach is expected to be valid for objects selected via a proxy for halo mass; however
if the selection criteria are sensitive to merger history (e.g. one selects active galactic nuclei)
then the universal mass function may not be valid. This is analogous to the merger bias
effect in models with primordial non-Gaussianity \cite{SlosarEtal, ReidEtal}.
 
In this picture, one assumes 
that galaxies form inside density peaks in Lagrangian space 
whose height exceeds some critical matter density contrast $\delta_c$
(i.e., $\rho > (1+\d_c) \rhob$).  In regions with
large-scale overdensity $\delta_l$, this threshold is effectively
lowered to $\delta_c-\delta_l$.  If we denote the average abundance
of tracers of mass $M$ as $\bar n(M,\d_c$), the galaxy density contrast 
on large scales is then linearly related to the matter density contrast 
via
\begin{align}
\delta_g(M;\delta_l)
= 
\left(1- \frac{\partial\ln\bar{n}(M,\delta_c,\tau)}
{\partial \delta_c}\right)\delta_l
= b \delta_l,\label{eq:bPBS}
\end{align}
if we truncate the Taylor expansion at linear order.  The first term
comes from mass conservation when transforming from Lagrangian to Eulerian
space.
As shown in the previous section, this argument is not in general
correct in the context of general relativity.  Note that if we
were to choose a non-synchronous gauge (such as conformal-Newtonian
or uniform-redshift gauge), the density field in different regions would be at 
different evolutionary stages,  so that the collapse threshold $\d_c$ is 
not simply a constant on a constant-coordinate-time hypersurface.  
Therefore, the galaxy density contrast must also depend on the evolutionary 
stage, or age of the universe, in the region considered.  
However, in a synchronized gauge where $\d\tau=0$, the argument leading
to \refeq{bPBS} is applicable.  Thus, \refeq{bPBS} is a valid bias
parameter which can be used in the correct, gauge-invariant bias expansion
\refeq{bias1}.   

We can also obtain a useful analytical estimate for $\bt$, assuming
that the abundance of galaxies follows a universal mass function,
\begin{equation}
\bar{n}(M) = \frac{\rhob}{M^2} f(\nu)
\left|\frac{d\ln\sigma}{d\ln M}\right|.
\end{equation}
First, the linear density bias is given by [\refeq{bPBS}]
\begin{equation}
b = 1 -\frac{\partial \ln \bar{n}}{\partial \delta_c}
= 1 -\frac{d \ln f(\nu)}{d \nu}\frac{1}{\sigma}.  
\end{equation}
On the other hand, $\bt$ is given by
\begin{align}
\bt =\:& \frac{\partial\ln (a^3\bar{n})}{\partial\ln a}
= \frac{\partial\ln f(\nu)}{\partial\ln a}
= \frac{d\ln f(\nu)}{d\nu} \frac{d\nu}{d\sigma} \frac{d\sigma}{d\ln a}
\nonumber
\\
=\:& (1-b) \sigma \left(-\frac{\nu}{\sigma}\right) \sigma \frac{d\ln\sigma}{d\ln a}.
\end{align}
The logarithmic derivative of $\sigma$ can be further simplified via
linear perturbation theory as
\begin{equation}
\frac{d\ln\sigma}{d\ln a} = \frac{d\ln D}{d\ln a} \equiv f.
\end{equation}
Here $f\approx \Omega_{\rm m}^{0.6}$ is the usual logarithmic growth rate familiar from redshift-space distortion theory \cite{kaiser:1987}.
In summary, we obtain
\begin{equation}
\bt = 
(b-1)\sigma\nu f
=
\delta_c f (b-1).
\label{eq:btauU}
\end{equation}
Note that the abundance of rarer, more strongly biased halos evolves
faster (larger $\bt$), and that the overall rate is set by the 
growth rate $f$.  \refeq{btauU}
is useful for estimating the magnitude of the corrections to the
galaxy power spectrum.  Note however that the universal mass function
prescription might not be a good description of actual tracers
whose redshift evolution is influenced by non-gravitational physics
(such as star formation, feedback, reionization, etc).  The key point
however is that for any given survey, $\bt$ is in fact observable,
if the redshift-dependence of the source selection function is known.  

%%%%%%%%%%%%%%%%%%%%%%%%%%%%%%%%%%%%%%%%%%%%%%%%%%%%%%%%%%%%%%%%%%%%%%%%%%%%%
\subsection{Bias in synchronous-comoving gauge}

In the synchronous-comoving gauge assumed in our derivation in \refsec{ng}, 
there is no perturbation to the
$00$-component of the metric.  Moreover, the constant-time hypersurfaces
are orthogonal to the velocities of comoving observers (in other
words, $v=0$).  In this gauge, every comoving observers' proper time is 
synchronized, and all observers on a given $\tau=$constant hypersurface
are at the same evolutionary stage, which implies $\d\tau=0$.  
The density field in standard $N$-body simulations is also defined
precisely in this gauge \cite{ChisariZaldarriaga}.  

We now see that the gauge-invariant bias relation \refeq{bias1}
is equivalent to the well-known linear bias relation between 
$\d_g$ and $\d_m$ in synchronous gauge,
\begin{equation}
\delta_g^\mathrm{(sc)} = b \delta_m^\mathrm{(sc)},
\end{equation}
where the superscripts denote that the variables are defined in
synchronous-comoving gauge.  Inserting this into \refeq{dgsc} then
yields the observed galaxy overdensity, completely described by
the metric and matter perturbations and two numbers specific to the
tracer population: the linear bias $b$ and the count slope $\bt$.  
Note that the latter parameter is observable in
galaxy surveys, while the bias $b$ is a parameter that needs to
be fitted for.    

%%%%%%%%%%%%%%%%%%%%%%%%%%%%%%%%%%%%%%%%%%%%%%%%%%%%%%%%%%%%%%%%%%%%%%%%%%%%%
\subsection{Primordial non-Gaussianity}\label{sec:nG}
In the picture outlined in this section, 
it is also straightforward to understand the
effect of primordial non-Gaussianity.  Consider a constant-age
hypersurface as defined earlier in this section, 
at some early time long before the
tracers of interest formed.  Since the
linear growth factor is the same everywhere on this slice, the
variance of the small-scale density field $\s_R^2$ smoothed on some scale $R$
is also the same everywhere, in the case of Gaussian initial conditions.  
In the presence of non-Gaussianity of the local type, mode-coupling induces
a modulation of $\s^2_R$ by long-wavelength (Bardeen) potential perturbations $\Phi_L$,
so that, within a region on a $t_U = $~const hypersurface 
where $\Phi_L$ can be considered constant, it is given by
\be
\hat\s_R^2 = \s_R^2 (1 + 4 f_{\rm NL} \Phi_L).
\ee
Here, $\s_R^2$ is the variance derived from the Gaussian part of $\Phi$.  
We see that this is closely related to perturbing the local age of 
the Universe, which leads to a change in the local $\s_R^2$ as well.  
The relation between $\Phi$ and $\d_m$ in synchronous-comoving gauge
is given by (e.g., \cite{DJS})
\begin{eqnarray}
\d_m^{\rm (sc)}(k,z) &=& \M(k,z) \Phi(k,z_*)
\nonumber \\ &=& \frac{3 \Omega_m H_0^2 k^2 (1+z)}{2 T(k) D(z)} \Phi(k,z_*),
\end{eqnarray}
where $z_*$ is some reference redshift where the non-Gaussian parameter
$f_{\rm NL}$ is defined (for example, that of the last-scattering surface).  
Using the universal mass function prescription (\refsec{univ}), we then see 
that the bias relation \refeq{bias1} in synchronous-comoving gauge is
modified to 
\be
\d_g^{\rm (sc)} = \left[b + 2 f_{\rm NL} (b-1)\d_c \M^{-1}(k)\right] \d_m^{(\rm sc)},
\ee
in agreement with \cite{DalalEtal08}.  
It is straightforward to generalize this derivation to more general
types of non-Gaussianity \cite{fsmk,DJS}.  Note in particular that
(for the local case), the scale-dependent correction to $\d_g^{\rm (sc)}$
is proportional to $k^{-2}$ out to arbitrarily large scales 
(see also \cite{WandsSlosar,BruniEtal}).

%%%%%%%%%%%%%%%%%%%%%%%%%%%%%%%%%%%%%%%%%%%%%%%%%%%%%%%%%%%%%%%%%%%%%%%%%%%%%
%%%%%%%%%%%%%%%%%%%%%%%%%%%%%%%%%%%%%%%%%%%%%%%%%%%%%%%%%%%%%%%%%%%%%%%%%%%%%
\section{The large-scale galaxy power spectrum}
\label{sec:Pkg}

We now calculate the observed galaxy power spectrum including the bias 
relation and the volume effect we have calculated in the previous sections.
Throughout this section, we use the cosmological parameters from Table 1
(``WMAP+BAO+$H_0$ ML'') of \citet{wmap7} as our reference cosmology.  
As explained in \refapp{derivdg}, we neglect the lensing contribution
$\k$, as it is not simply incorporated into a three-dimensional
power spectrum.  Further, we neglect two very small contributions,
the integrated Sachs-Wolfe (ISW) contribution to $\d z$, and the 
time-delay contribution $\propto \int d\chi\: (\phi+E'')/\chit$.  
We also neglect the stochastic contribution $\eps$ to $\d_g$ in the 
following, and assume that the primordial density perturbation follows 
Gaussian statistics.

Neglecting the ISW contribution, we have (\refapp{ISW})
\be
\delta z(\chit) = \partial_\parallel E'(\chit) + E''(\chit),
\ee
where we have dropped the unobservable, constant contribution from
the perturbations evaluated at $o$.  Using \refeq{dgsc}, \refeq{magsc}, and the results 
from \refsec{bias}, the observed galaxy density contrast written in terms
of perturbations in synchronous-comoving gauge is then given by
(see \refapp{Pgderiv} for the derivation)
\ba
\tilde\delta_g
=\:&
b\delta_m^{\rm (sc)} 
+\bt(\partial_\parallel E'+E'')
+2(1 - \Q)\phi 
-\frac{\partial^2_\parallel E'}{aH}
\nonumber
\\
&
-\frac{2}{\tilde\chi}(1-\Q) E' 
+(\mathcal{C}-1)(\partial_\parallel E'+E''),
\label{eq:dgreal}
\ea
where
\begin{equation}
\mathcal{C} = \frac{1+z}{H}\frac{dH}{dz} 
- \frac{1+z}{H}\frac{2}{\tilde\chi} (1 - \Q) - 2 \Q.
\label{eq:def_C}
\end{equation}
In a $\Lambda$CDM Universe, the first term in $\mathcal{C}$ can be simplified to yield
\begin{equation}
\mathcal{C} 
= 
\frac{3}{2}\Omega_m(z)
- \frac{1+z}{H}\frac{2}{\tilde\chi}(1 - \Q) - 2 \Q.
\end{equation}
where $\Omega_m(z)$ is the matter density parameter at redshift $z$.  
In Fourier space, \refeq{dgreal} reads
\ba
\tilde\delta_g(\bm{k})
=\: & b \delta_m^{(sc)} 
+\bt
( ik\mu E'+E'' )
  + 2(1-\Q)\phi 
   +\mu^2\frac{k^2 E'}{aH}
   \vs
&
	-(1-\Q)\frac{2E'}{\tilde\chi} 
	+(\mathcal{C}-1) ( ik\mu E'+E'' ), 
\label{eq:dgFourier}
\ea
where $\mu$ is the cosine of the wave-vector $\bm{k}$ 
with the line-of-sight direction.  

% !!!!!!!!!!!!!!!!!!!!!!!!!!!
\begin{figure}
\centering
\includegraphics*[width=0.48\textwidth]{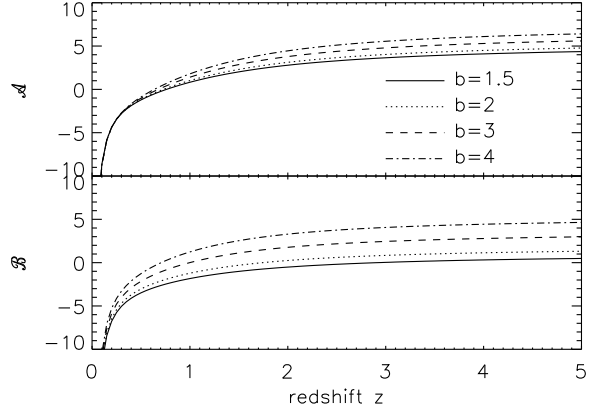}
\caption{%
Coefficients of new terms in observed galaxy overdensity 
in synchronous comoving gauge \refeq{dgk}
introduced by relativistic volume effect and bias.
Here, we plot for $b=1.5$ (solid), $b=2$ (dotted), $b=3$ (dashed)
and $b=4$ (dot-dashed) cases, and $\bt$ is calculated assuming the 
universality of the mass function [\refeq{btauU}].
We ignore the magnification effect by setting $\Q=0$.
}%
\label{fig:coeffs}
\end{figure}
% !!!!!!!!!!!!!!!!!!!!!!!!!!!

Finally, by relating the synchronous-comoving gauge metric perturbations to 
the density contrast $\d_m$ (\refapp{sc_variables}), we can further simplify 
$\tilde \d_g$ as
\begin{align}
\frac{\tilde\delta_g}{\delta_m}
=\:& b+f\mu^2
+ \frac{\mathcal{A}}{x^2}
+ \frac{i\mu}{x}\mathcal{B}
\label{eq:dgk},
\end{align}
where $x\equiv k/aH$ is the wavenumber in units of the comoving horizon,
and we have defined the coefficients 
\ba
\mathcal{A} =\frac{3}{2}\Omega_m \Bigg[ &
\bt\left(1-\frac{2f}{3\Omega_m}\right)
+1+\frac{2f}{\Om}
+\mathcal{C}-f - 2 \Q
\Bigg]
\ea
and
\be
\mathcal{B} =
f\left[\bt+\mathcal{C}-1 \right].
\ee
For given value of $b_e$ and $\Q$, $\mathcal{A}$ and $\mathcal{B}$ 
are only functions of redshift, and incorporate the relativistic bias
as well as the volume distortion and magnification effects.  
On small scales, when $x \gg 1$, we recover the usual Fourier-space
galaxy overdensity in ``Newtonian theory'',
\begin{equation}
\tilde\delta_g \stackrel{x \gg 1}{=} (b+f\mu^2)\delta_m.
\end{equation}

In principle, $\bt$ can be measured from the survey itself provided 
one has good knowledge of the redshift-dependence of the selection
function.  In the following, we will use
\be
\bt = \delta_{\rm c} f(b-1)
\ee
as predicted by the universal mass function ansatz [\refsec{univ}] for our
illustrations.  
Figure~\ref{fig:coeffs} shows $\mathcal{A}$ and $\mathcal{B}$ as function
of redshift when we only include relativistic volume and bias effect, i.e. 
$\Q=0$.
Note that both $\mathcal{A}$ and $\mathcal{B}$ diverge to $-\infty$ as
$z\to 0$, because of the $1/\chit$ factor in $\mathcal{C}$ [\refeq{def_C}]
which is a result of the volume distortion by velocities and gravitational
redshifts.  
Figure~\ref{fig:coeffs2} shows $\mathcal{A}$ and $\mathcal{B}$ for fixed
bias ($b=2$), and varying magnification coefficient 
($\Q=-1$, $0.5$, $1$, and $2$).  
For $\Q=1$, which is the case for diffuse backgrounds,
magnification effect cancels volume distortion, and both $\mathcal{A}$
and $\mathcal{B}$ are small.
For sufficiently high redshift when the Universe is approximately matter 
dominated, 
$\mathcal{C}\to 3\Om/2-2\Q\simeq 1.5-2\Q$ and $f\to 1$, so that 
$b_e\to \delta_c(b-1)$ and we can approximate the coefficients as
$\mathcal{A}\to 5.2 -6\Q +\delta_c(b-1)/2 $,
$\mathcal{B}\to 0.5 -2\Q +\delta_c(b-1)$. 

% !!!!!!!!!!!!!!!!!!!!!!!!!!!
\begin{figure}
\centering
\includegraphics*[width=0.48\textwidth]{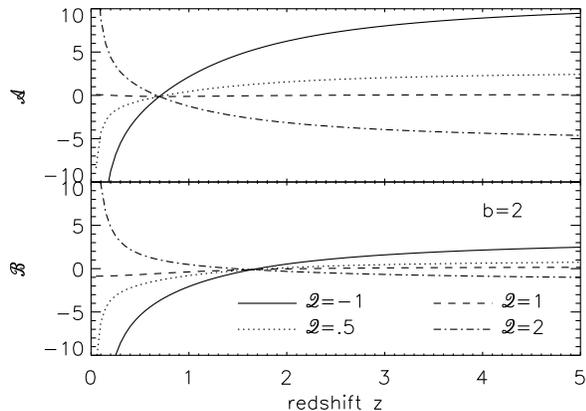}
\caption{%
Same as \reffig{coeffs}, but for fixed bias ($b=2$) and varying
magnification with $\Q=-1$ (solid), $0.5$ (dotted), $1$ (dashed), 
$2$ (dot-dashed). For diffuse backgrounds, $\Q=1$, and the magnification effect
cancels almost all volume distortions.
}%
\label{fig:coeffs2}
\end{figure}
% !!!!!!!!!!!!!!!!!!!!!!!!!!!

From \refeq{dgk} we can calculate the observed galaxy
power spectrum in terms of the linear matter power spectrum $P^{(\rm sc)}(k)$
in synchronous comoving gauge, yielding
\be
\frac{P_g(k,\mu)}{P^{(\rm sc)}(k)}
=
(b+f\mu^2)^2 + 2(b+f\mu^2)\frac{\mathcal{A}}{x^2} + \frac{\mathcal{A}^2}{x^4}
+
\frac{\mu^2\mathcal{B}^2}{x^2}.
\label{eq:Pkg}
\ee
Note again that we have neglected all projected quantities here,
most importantly the magnification contribution $-2\k$.  
Furthermore, the flat-sky calculation employed here is likely not applicable
to the very largest scales in actual galaxy surveys.  
Here we are mainly interested in the issue of galaxy biasing however,
and defer the calculation of the full angular galaxy power spectrum to future work.  

% !!!!!!!!!!!!!!!!!!!!!!!!!!!
\begin{figure}[t!]
\centering
\rotatebox{90}{
\includegraphics*[height=0.48\textwidth]{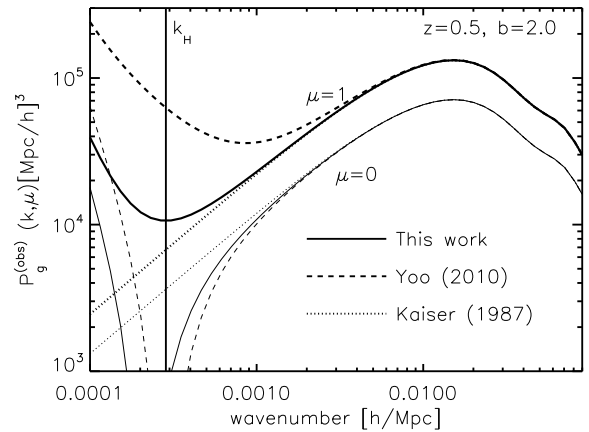}
}
\caption{
Three different theoretical predictions of the observed galaxy power spectrum on
large scales for galaxies with linear bias $b=2$ at redshift $z=0.5$,
and assuming $\Q=0$:
Newtonian linear theory \cite{kaiser:1987} (dotted line),
relativistic linear theory with linear bias in uniform redshift gauge 
\cite{yoo/etal:2009} (dashed line), and 
relativistic linear theory with linear bias in synchronous comoving gauge 
(this work, solid line).
We show both line-of-sight directional power spectrum ($\mu=1$, thick lines)
and perpendicular directional power spectrum ($\mu=0$, thin lines).  
The vertical solid line indicates $k = aH$ at $z=0.5$.  
}
\label{fig:Pk-comp}
\end{figure}
% !!!!!!!!!!!!!!!!!!!!!!!!!!!

\reffig{Pk-comp} shows the galaxy power spectrum in three different
calculations, each for $\mu=0$ and $\mu=1$.  The dotted lines show
the linear small-scale limit given by the Kaiser formula.  The solid
lines show the relativistic calculation \refeq{Pkg} using the galaxy bias
prescription derived in \refsec{bias}, i.e. a linear bias relation
in synchronous-comoving gauge.  Clearly,
the prediction departs from the Kaiser formula on scales $k\lesssim 10^{-3}\iMpch$.  
The dashed lines in \reffig{Pk-comp} show the result of \cite{yoo/etal:2009}
for comparison.  As we have seen in \refsec{bias}, their result is
equivalent to linear biasing in the uniform-redshift gauge.  We
see that the departures from the small-scale limit are much more significant
in this latter calculation, showing that the precise choice of bias relation 
is important on very large scales.  

\reffig{Pk2D} shows the two-dimensional galaxy power spectrum from
\refeq{Pkg}, illustrating the evolution of the angular dependence of
$P_g(k,\mu)$ with scale.  Also shown for comparison is the prediction
of the Kaiser formula \cite{kaiser:1987}.  
Again, deviations appear for $k\lesssim 10^{-3}\iMpch$
and become more significant for transverse separations.  

% !!!!!!!!!!!!!!!!!!!!!!!!!!!
\begin{figure*}
\centering
\rotatebox{90}{
\includegraphics*[width=0.7\textwidth]{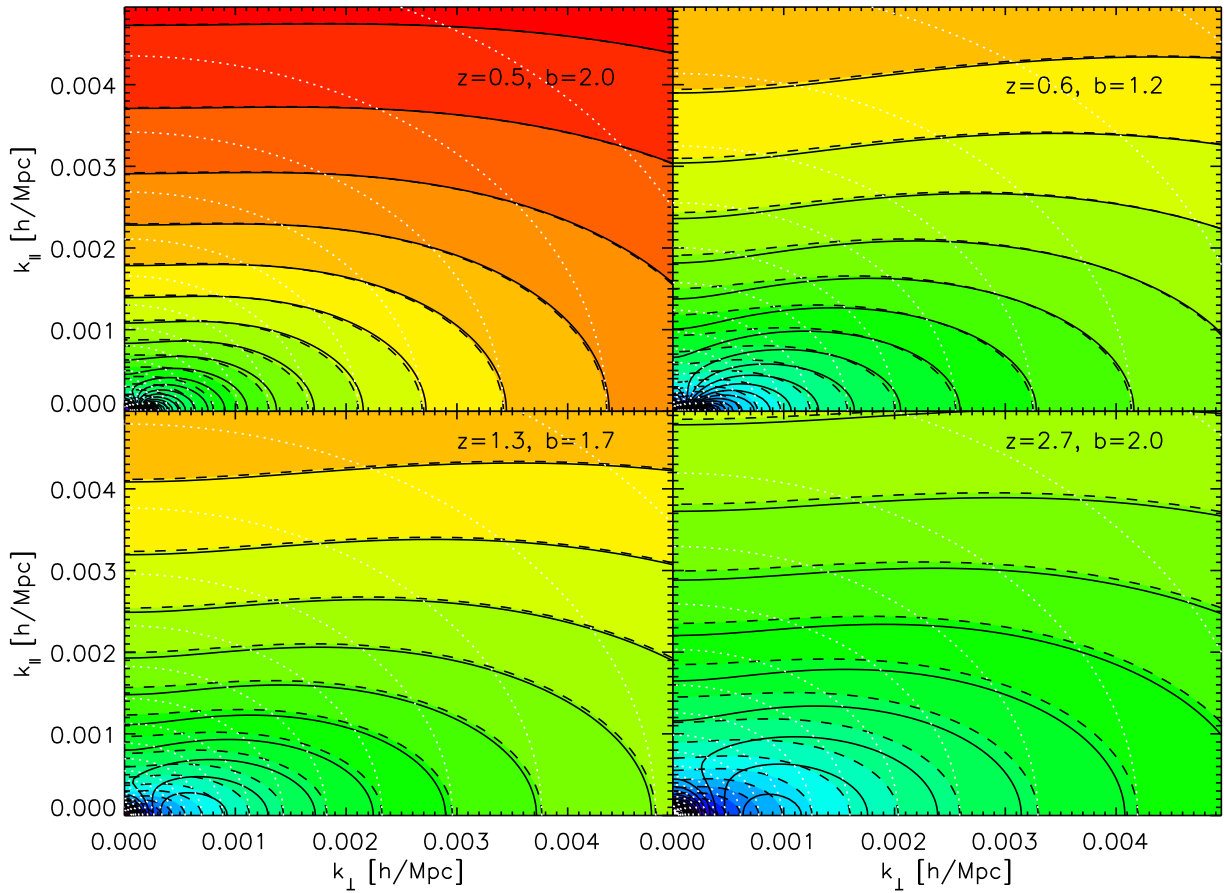}
}
\caption{
Observed galaxy power spectrum (\refeq{Pkg}) as function
of $k_{\parallel}$ and $k_\perp$.  
The color contours and dashed lines 
show the Newtonian result from Kaiser \cite{kaiser:1987}, 
while the solid lines show the result including relativistic corrections.  
Here we have set $\Q=0$.
For reference, we also show the real space power 
spectrum contours as white dotted lines.
}%
\label{fig:Pk2D}
\end{figure*}
% !!!!!!!!!!!!!!!!!!!!!!!!!!!

%%%%%%%%%%%%%%%%%%%%%%%%%%%%%%%%%%%%%%%%%%%%%%%%%%%%%%%%%%%%%%%%%%%%%%%%%
\subsection{Effective primordial non-Gaussianity}

Our results show that the observed power spectrum departs from the
small-scale calculation on sufficiently large scales, with terms
proportional to $(k/a H)^{-2}$ and $(k/aH)^{-4}$.  This is reminiscent
of the scale-dependent bias induced by primordial non-Gaussianity
\cite{DalalEtal08}, see \refsec{nG}.  
It is thus natural to compare the two effects.  

Neglecting all relativistic effects, and non-Gaussian redshift distortion terms
that are unimportant on large scales \cite{Schmidt10},
the observed galaxy power spectrum for local 
primordial non-Gaussianity is given by
\ba
\frac{P_g^{(\rm NG)}(k)}{P^{(\rm sc)}(k)} =\:& (b + \D b(k) + f \mu^2)^2 
\label{eq:PkNG}\\
=\:& (b+f\mu^2)^2 + 2 \D b(k) (b+f\mu^2) + \D b^2(k).\nonumber
\ea
Here we have defined the bias correction
\ba
\Delta b(k) =\:& \frac{2\fnl(b-1)\d_{\rm c}\,3\Omega_{m0}H_0^2}{2D(z)k^2T(k)}\vs
=\:& \fnl\delta_{\rm c}(b-1) \frac{3\Omega_{m}(z) H^2a^3}{D(z)k^2T(k)}\vs
\simeq\:& \frac{3}{2x^2}\Omega_m(z) \delta_{\rm c}(b-1) \fnl \frac{2a}{D(z)},
\ea
where in the last line we have assumed that $k \lesssim 0.01\iMpch$ 
so that $T(k) = 1$.  We can define an effective
non-linearity parameter $\fnl^{\rm eff}$ such that the scale-dependent
non-Gaussian bias inserted into the small-scale expression \refeq{PkNG} 
leads to a power spectrum matching the actual observed power spectrum
in the Gaussian case including the relativistic effects.  
However, the relativistic corrections affect modes perpendicular 
and parallel to the line of sight differently, while the non-Gaussian
scale-dependent bias is isotropic. [Note in particular the $\mathcal{B}$ term in Eq.~(\ref{eq:Pkg}).]
Therefore, we can, in principle, distinguish the relativistic 
effect from the non-Gaussian scale dependent bias through the angular dependence
of the 2D power spectrum.   Given the limited number of modes on scales
$k\lesssim 10^{-3} h/\rm Mpc$, it is more realistic to consider
the effect on the angle-averaged power spectrum monopole
\be
P^{\rm (\ell=0)}_g(k) = \frac{1}{2}\int_{-1}^1 d\mu\: P_g(k,\mu).
\label{eq:Pkmono}
\ee
From \refeq{Pkg}, we find for the relativistic corrections to the 
monopole power spectrum
\ba
&\frac{
P^{\rm (\ell=0)}_g(k)
-
P^{\rm (Kaiser,\ell=0)}_g(k)
}{P^{\rm (sc)}(k)}\vs
&{\rm~~~~}=
\left[
2\left(b+\frac{f}{3}\right)\mathcal{A}
+
\frac{\mathcal{B}^2}{3}
\right]
\frac{1}{x^2}
+ \frac{\mathcal{A}^2}{x^4},
\label{eq:Pkmono_GR}
\ea
and from \refeq{PkNG}, we find that the non-Gaussian corrections only
are given by
\ba
&\frac{
P^{\rm (NG, \ell=0)}_g(k)
-
P^{\rm (Kaiser,\ell=0)}_g(k)
}{P^{\rm (sc)}(k)}\vs
&{\rm~~~~}=
\frac{3}{x^2}\Omega_m \delta_c(b-1) \fnl \frac{2a}{D(z)}
\left[
b+\frac{f}{3}
\right]
+
\Delta b^2(k),
\label{eq:Pkmono_NG}
\ea
where 
\be
P^{\rm (Kaiser,\ell=0)}_g(k)
=
\left[
b^2 + \frac{2}{3}bf + \frac{1}{5}f^2
\right]
P_m^{\rm (sc)}(k)
\ee
is the monopole galaxy power spectrum given by the Kaiser formula.
By equating the term proportional to $x^{-2}$ in
\refeq{Pkmono_GR} and \refeq{Pkmono_NG}, we find the effective 
amplitude of non-Gaussianity as
\ba
\fnl^{\rm eff}
=\:&
\frac{D(a)}{a}
\frac{1}{6\Om\delta_c(b-1)}
\left[
2\mathcal{A} + \frac{\mathcal{B}^2}{3b+f}
\right]
\label{eq:efffnl}\\
=\:&
\frac{1}{2}
\frac{D(a)}{a}
\biggl[
f\left(1-\frac{2f}{3\Om}\right)
\vs
&\qquad\quad+
\frac{1+2f/\Om+\mathcal{C}-f-2\Q}{\delta_c(b-1)}
\vs
&\qquad\quad+ 
\frac{\delta_c(b-1)}{3\Om(3b+f)}
\left(
f^2+f\frac{\mathcal{C}-1}{\delta_c(b-1)}
\right)^2
\biggl].
\nonumber
\ea
\reffig{PkNG} shows the monopole of the galaxy power spectrum,
using the full expression \refeq{Pkg} and \refeq{PkNG} with \refeq{efffnl}.  
They generally agree very well, apart from the lowest-redshift case.  

% !!!!!!!!!!!!!!!!!!!!!!!!!!!
\begin{figure*}
\centering
\rotatebox{90}{
\includegraphics*[width=0.7\textwidth]{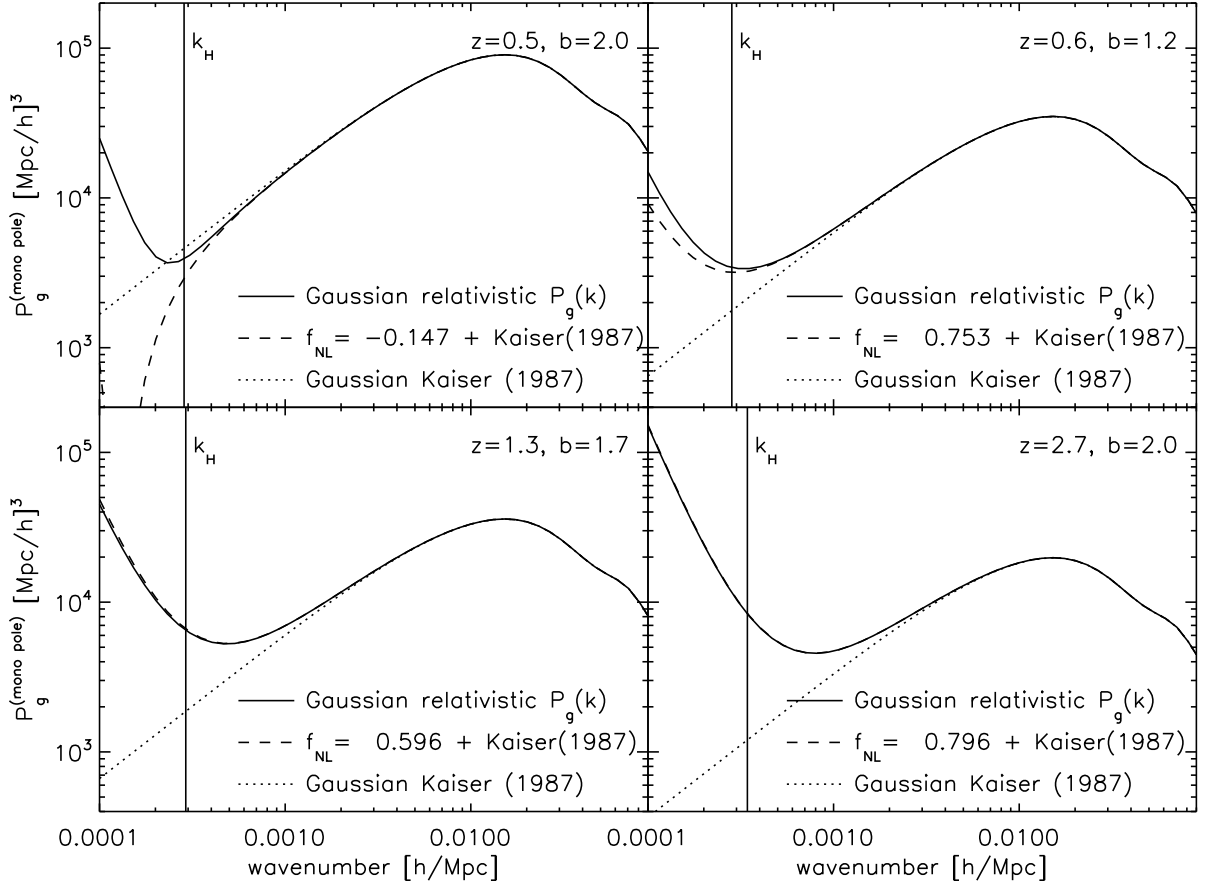}
}
\caption{
Monopole galaxy power spectrum [\refeq{Pkmono}] from the full
expression \refeq{Pkg}, and using the effective $\fnl$ approximation,
\refeq{PkNG}.  
We use the same set of bias and redshift as in \reffig{Pk2D} (and $\Q=0$).
The effective $\fnl$ approximation is valid
within the horizon ($k > a H$), marked by a vertical line, 
while it breaks down on large scales where the
$x^{-4}$ term in \refeq{Pkmono_GR} is important.  
The breakdown is more apparent when $\mathcal{B}$ is large
(upper two panels).   
}
\label{fig:PkNG}
\end{figure*}
% !!!!!!!!!!!!!!!!!!!!!!!!!!!

The effective $\fnl$ given by \refeq{efffnl} can serve as a useful
tool to forecast for a given survey whether relativistic effects become
important:  if a survey achieves a forecasted precision on the local $\fnl$
of order $\fnl^{\rm eff}$, then relativistic effects are relevant.  
\reffig{efffnl} shows $\fnl^{\rm eff}$ for differently biased tracers
as a function of redshift when $\Q=0$.  Typical values are around 0.2, and
generally lower than 0.5.  Note that we have assumed the universal
mass function approach to estimate $\fnl^{\rm eff}$;  if for some reason
the tracer number density evolves very rapidly with redshift so that
$\bt$ becomes large, $\fnl^{\rm eff}$ would increase
correspondingly.  In general however, we expect that number to
remain less than 1.  
$\fnl^{\rm eff}$ rises sharply for $z\rightarrow 0$, because of 
the $1/\chit$ factor in $\mathcal{C}$ [\refeq{def_C}].  
\reffig{efffnl2} shows $\fnl^{\rm eff}$ for galaxies of bias $b=2$ 
with different luminosity function with slope of ($\Q=-1$, $0.5$, $1$, $2$).  
This figure indicates that $\fnl^{\rm eff}$ varies with $\Q$, but does 
not exceed 2 except for low redshifts
where, again, $\fnl^{\rm eff}$ diverges due to $1/\chit$.
This divergence however, is removed completely for diffuse backgrounds
where $\Q=1$ (dashed line).

% !!!!!!!!!!!!!!!!!!!!!!!!!!!
\begin{figure}
\centering
\includegraphics*[width=0.48\textwidth]{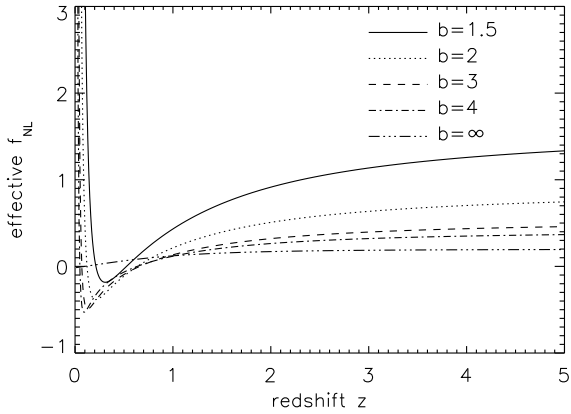}
\caption{Effective $f_{\rm NL}$ [\refeq{efffnl}] as a function
of redshift for different galaxy biases when $\Q=0$.}
\label{fig:efffnl}
\end{figure}
% !!!!!!!!!!!!!!!!!!!!!!!!!!!

% !!!!!!!!!!!!!!!!!!!!!!!!!!!
\begin{figure}
\centering
\includegraphics*[width=0.48\textwidth]{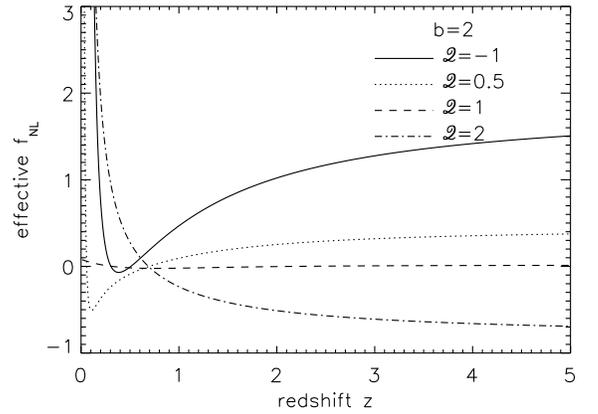}
\caption{
Same as \reffig{efffnl}, but 
for galaxy bias ($b=2$) with different magnification 
$\Q=-1$, $0.5$, $1$ and $2$.}
\label{fig:efffnl2}
\end{figure}
% !!!!!!!!!!!!!!!!!!!!!!!!!!!

%%%%%%%%%%%%%%%%%%%%%%%%%%%%%%%%%%%%%%%%%%%%%%%%%%%%%%%%%%%%%%%%%%%%%%%%%
%%%%%%%%%%%%%%%%%%%%%%%%%%%%%%%%%%%%%%%%%%%%%%%%%%%%%%%%%%%%%%%%%%%%%%%%%
\section{Conclusions}
\label{sec:concl}

Future galaxy surveys will measure the clustering of galaxies on
scales approaching the horizon.  Thus, it is necessary to embed 
the observed galaxy density in a relativistic context.  This 
problem has received considerable attention recently.  In this
paper, we have derived the observed galaxy density contrast $\tilde\d_g$
in terms of the intrinsic galaxy overdensity $\d_g$ and metric
perturbations in the synchronous-comoving gauge (\refsec{ng}).  
By transforming to conformal-Newtonian gauge, we reach
agreement with the results of Refs.~\cite{challinor/lewis:2011,bonvin/durrer:2011}.  
On the other hand, we find some minor disagreement with the expression 
of \cite{yoo/etal:2009}, which can be traced back to a sign issue.  
In \refapp{test}, we also show that our formula for $\tilde\d_g$ 
reproduces the correct analytic result for the following five test cases:
1) pure spatial gauge mode, 2) zero-wavenumber gauge mode,
3) perturbation to the expansion history,
4) spatial curvature, 5) Bianchi I cosmology.  These tests cover most of 
the parameter space of actual cosmologies in a simplified setting, and 
thus lend credibility to our result.  

One necessary, further ingredient for a description of galaxy clustering
on large scales is a physical, gauge-invariant definition of galaxy
bias.  We present a straightforward physically motivated definition
in \refsec{bias}.  This bias relation is easily seen to reduce to the
standard linear bias relation in synchronous coordinates, where 
constant-time hypersurfaces coincide with constant-age hypersurfaces.  
The bias relation \refeq{bias1} can be seen as a proper generalization
of the peak-background split bias.  Using this result, we arrive
at a simple expression for the observed galaxy density perturbations
in synchronous-comoving gauge [\refeq{dgreal}], described by the
(gauge-invariant) bias parameter $b$, and the redshift evolution
of the tracer population [through $b_e$, \refeq{btaudef}].  

The recent study of \citet{BaldaufEtal} has also reached the same conclusion by
constructing locally flat space-time coordinates around a freely-falling
observer, where long-wavelength modes locally act as curvature.  
In that coordinate system, the galaxy number density is modulated not only 
by the local curvature but also the time difference between global time and 
local time whose effect exactly coincides with our $\bt$ in \refeq{bias1}.  
In contrast, Ref.~\cite{yoo/etal:2009} adopted a bias relation
in the constant-observed-redshift gauge, which leads to considerably
different predictions for the large-scale galaxy power spectrum.  
However, this does not seem to be a physical description of galaxy bias,
since the age of the Universe is \emph{not} constant on 
constant-observed-redshift slices.

We believe these results, together with other recent work
\cite{challinor/lewis:2011,bonvin/durrer:2011,BaldaufEtal}, allow
us to unambiguously predict the two-point statistics of large-scale
structure tracers on large scales.

%%%%%%%%%%%%%%%%%%%%%%%%%%%%%%%%%%%%%%%%%%%%%%%%%%%%%%%%%%%%%%%%%%%%%%%%
\acknowledgments
We would like to thank Olivier Dor\'e and the participants of the
Michigan non-Gaussianity workshop for helpful discussions.
FS and DJ are supported by, respectively, 
the Gordon and Betty Moore Foundation and 
Robinson prize postdoctoral fellowship at Caltech.
%%%%%%%%%%%%%%%%%%%%%%%%%%%%%%%%%%%%%%%%%%%%%%%%%%%%%%%%%%%%%%%%%%%%%%%%
%%%%%%%%%%%%%%%%%%%%%%%%%%%%%%%%%%%%%%%%%%%%%%%%%%%%%%%%%%%%%%%%%%%%%%%%
\appendix

%%%%%%%%%%%%%%%%%%%%%%%%%%%%%%%%%%%%%%%%%%%%%%%%%%%%%%%%%%%%%%%%%%%%%%%%%%%%%
\section{Metric variables and gauge transformations}
\label{app:gTransf}

Let us consider a general scalar coordinate transformation ($T$, $L$)
%from old coordinate ($x^\alpha$) to new coordinate ($\check{x}^\alpha$):
\begin{equation}
\label{eq:cTrans}
x^\alpha\to
\check{x}^\alpha = x^\alpha + \left(T(x^\alpha),\partial^i L(x^\alpha) \right).
\end{equation}
While true scalar quantities are invariant under such a coordinate change,
perturbations around the background do change because the background
is time-dependent.

For example, consider a scalar function $\zeta(\tau,\vx)$ whose 
background value only depends on time $\bar\zeta(\tau)$. This is the case for
all scalar functions in the homogeneous universe.
As a scalar, the function $\zeta$ does not change
under the coordinate transformation in \refeq{cTrans}, 
and it is only the background time $\tau$, thus $\bar\zeta(\tau)$, that is changed.
From the relation
\ba
\zeta 
= \bar\zeta(\tau)  + \delta \zeta(\tau,\vx)
= \bar\zeta(\check\tau)  + \widecheck{\delta \zeta}(\check\tau,\check\vx),
\ea
we calculate the scalar perturbation $\widecheck{\delta\zeta}$ in the
transformed coordinate in terms of the variables in the old coordinate as
\be
\widecheck{\delta\zeta}
=
\delta\zeta + \bar\zeta(\tau) - \bar\zeta(\tau+T)
=
\delta\zeta - \frac{d\bar\zeta(\tau)}{d\tau}T 
\label{eq:gTscalar}
\ee
up to linear order in perturbations. 
By appyling \refeq{gTscalar}, we find that the matter density contrast 
and redshift perturbation transform as
\begin{align}
\check{\delta}_m 
=\:&
\delta_m - \frac{d\ln\rho_m}{d\tau} T
=
\delta_m + 3 aHT,\nonumber\\
\widecheck{\delta z}
=\:&
\delta z+aHT,{\rm~~and}\nonumber\\
\check{\delta}_g 
=\:&
\delta_g - b_e aHT,
\label{eq:GTscalar}
\end{align}
where in the second line we have used that $d\bar z/d\tau = -aH$.  
Hence, the quantity $\delta_m-3\delta z$ remains invariant under 
gauge transformations.  The scalar component of the peculiar velocity, 
$\partial_i v \equiv adx^i/dt$, transforms as
\begin{equation}
\check{v} = v + L'.
\label{eq:GTvector}
\end{equation}
We can calculate the gauge transformation of 
higher rank vectors/tensors in a similar way, by applying the appropriate 
transformation law of the object.  

% % % % % % % % % % % % % % % % % % % % % % % % % % % % % % % % % 
\subsection{General scalar metric perturbations}

Consider a general FRW metric with scalar perturbations defined through
\begin{align}
\label{eq:metric_latin}
ds^2 =\:& -a^2 (1+2A) d\tau^2
-2a^2 B_{,i} d\tau dx^i
\nonumber
\\
&
+a^2 \left[
(1+2D)\delta_{ij}
+2E_{ij}
\right]dx^i dx^j
\end{align}
where
\be
E_{ij} = 
\left(\partial_i\partial_j - \frac{1}{3}\delta_{ij}\nabla^2 \right) E.
\label{eq:Escalar}  
\ee
Then the metric perturbations transform as follows: 
\begin{align}
\check{A} =\:& A - aH T - T'
,\nonumber\\
\check{B} =\:& B + L' - T
,\nonumber\\
\check{D} -\frac{1}{3}\nabla^2 \check{E}=\:& 
D -\frac{1}{3}\nabla^2 E - aHT
,{\rm~~and}\nonumber\\
\check{E} =\:& E -L.
\end{align}
Similarly, when the perturbed metric is defined through
\begin{align}
\label{eq:metric_greek}
ds^2 =\:& -a^2 (1+2\alpha) d\tau^2
-2a^2 \beta_{,i} d\tau dx^i
\nonumber
\\
&
+a^2 \left[
(1+2\varphi)\delta_{ij}
+2\gamma_{,ij}
\right]dx^i dx^j,
\end{align}
perturbations transform as follows: 
\begin{align}
\check{\alpha} =\:& \alpha - aH T - T'
,\nonumber\\
\check{\beta} =\:& \beta + L' - T
,\nonumber\\
\check{\varphi} =\:& 
\varphi - aHT
,{\rm~~and}\nonumber\\
\check{\gamma} =\:& \gamma -L.
\label{eq:GTmetric}
\end{align}
Note that the spatial metric components in \refeq{metric_latin} and
\refeq{metric_greek} are related through
\be
\alpha = A {\rm~~and~~} \varphi = D - \frac{1}{3}\nabla^2E.  
\ee

% % % % % % % % % % % % % % % % % % % % % % % % % % % % % % % % % 
\subsection{From synchronous-comoving to conformal-Newtonian gauge}
\label{app:sc2cN}

The two most commonly considered gauges in cosmology are the
conformal-Newtonian gauge, which is defined through
\be
B = E = 0 \quad\Leftrightarrow\quad \beta = \gamma = 0,
\ee
using the convention \refeq{metric_latin} and \refeq{metric_greek},
respectively, and the synchronous-comoving gauge, defined
through
\be
A = B = v = 0 \quad\Leftrightarrow\quad \alpha = \beta = v = 0,
\ee
in the same conventions.  Here we explicitly give the transformation between 
these two gauges, using the general expressions of the previous section.  
We find that
\begin{align}
0 =\:& \alpha - aH T - T'
,\nonumber\\
0 =\:&  L' - T
,\nonumber\\
D - \frac{1}{3} \nabla^2 E=\:& \varphi - aHT
,\nonumber\\
E =\:& -L
,{\rm~~and}\nonumber\\
0 =\:& v+ L'.
\end{align}
Solving these equations leads to the transformation from conformal-Newtonian
to synchronous-comoving gauge,
\be
D = \varphi + \frac{1}{3}\nabla^2 \int d\tau v + aHv
{\rm~~and~~}
E = \int d\tau v,
\label{eq:syncpg}
\ee
and the corresponding inverse transformation,
\begin{align}
\alpha =\:& - aHE' - E''
,\nonumber\\
\varphi =\:& 
D-\frac{1}{3}\nabla^2 E - aH E' 
,{\rm~~and}\nonumber\\
\label{eq:conav}
v =\:& E'.
\end{align}
Note that in \refeq{syncpg},
we have the freedom to add an integration constant, 
which reflects the fact that $D$ and $E$ in synchronous gauge 
contain a spatial gauge mode.   
Such a gauge mode can be removed by introducing 
$\phi\equiv D-\nabla^2 E/3$ and/or taking time derivatives 
of $E$.  For example, Eq.~(\ref{eq:conav}) contains only
$\phi$ and $E'$, because the conformal-Newtonian gauge 
has no residual gauge freedom.

% % % % % % % % % % % % % % % % % % % % % % % % % % % % % % % % % 
\subsection{Metric variables in synchronous gauge}
\label{app:sc_variables}

In this appendix, we show the relation among three common parametrizations
of synchronous-comoving gauge, and how to relate all metric perturbations
to the matter density perturbation in the adiabatic case.  

The convention used in this paper is
\be
\delta g_{ij}(\tau,\vx) = a^2(\tau)
\left[
2D(\tau,\vx)\delta_{ij} + 2 E_{ij}(\tau,\vx)
\right],
\label{eq:gij1}
\ee
where $E_{ij}$ is traceless and given in \refeq{Escalar}.  
In \citet{yoo:2010}, the spatial metric perturbation is defined as
\be
\delta g_{ij}(\tau,\vx) = a^2(\tau)
\left[
2\phi(\tau,\vx)\delta_{ij} + 2 \partial_i\partial_j\gamma(\tau,\vx)
\right].
\label{eq:gij2}
\ee
Comparing Eqs.~(\ref{eq:gij1},\ref{eq:gij2}), we find the relations
\be
\phi(\tau,\vx) = D(\tau,\vx) - \frac{1}{3}\nabla^2 E(\tau,\vx)
\,{\rm~and~}\, \gamma(\tau,\vx) = E(\tau,\vx).
\ee

In Fourier space, the metric in \refeq{gij1} becomes
\begin{eqnarray}
\delta g_{ij}(\tau,\vk) &=& a^2(\tau)
\Big[
2D(\tau,\vk)\delta_{ij}  \label{eq:gij1-1} \nonumber \\
&& - 2 \left(
\vk_i\vk_j - \frac{1}{3}k^2\delta_{ij}
\right)E(\tau,\vk)
\Big],
\end{eqnarray}
which can be compared with the spatial metric perturbations
defined in Ma \& Bertschinger \cite{ma/bertschinger:1995}:
\ba
\delta g_{ij}(\bm{k},\tau)
= \;&
a^2(\tau)
\Big[
\frac{\bm{k}_i\bm{k}_j}{k^2} h(\bm{k},\tau)
\label{eq:gij3} \nonumber\\
&+
6
\left(
\frac{\bm{k}_i\bm{k}_j}{k^2} 
-\frac{1}{3}\delta_{ij}
\right)
\eta(\bm{k},\tau)
\Big].
\ea
Comparing \refeq{gij1-1} and \refeq{gij3} leads to
\be
D(\tau,\vk) 
=
\frac{h(\tau,\vk)}6
{\rm~~and~~}
E(\tau,\vk) 
=
-\frac{h(\tau,\vk)+6\eta(\tau,\vk)}{2k^2},
\ee
and combining above results yields
\be
\phi(\tau,\vk) = -\eta(\tau,\vk)
{\rm~~and~~}
\gamma(\tau,\vk) = -\frac{h(\tau,\vk)+6\eta(\tau,\vk)}{2k^2}.
\ee

We now relate the metric variables to the matter density contrast in 
synchronous gauge.  These relations assume General Relativity, Gaussianity,
and adiabaticity, and are only used for the 
equatiosn and numerical results in \refsec{Pkg}.  
First, let us consider the time derivative of $E$.
The continuity equation in synchronous comoving
gauge is given by
\cite{ma/bertschinger:1995} 
 \begin{equation}
\delta_m'(\tau,\vk) = -\frac{1}{2}h'(\tau,\vk),
\end{equation}
and from the Einstein equations we have
[e.g. Eq.~(22) of \cite{ma/bertschinger:1995}]
\begin{equation}
k^2\eta'(\tau,\vk) = 4i \pi Ga^2 k^j\delta T^0_{j} \propto v = 0.
\label{eq:etaprime}
\end{equation}
Therefore, we calculate the time derivative of $E$ as
\begin{equation}
E'(\tau,\vk) 
= -\frac{h'(\tau,\vk)}{2k^2} 
= \frac{\delta_m'(\tau,\vk)}{k^2}
=
\frac{aHf}{k^2}\delta_m(\tau,\vk),
\end{equation}
where $f=d\ln D/d\ln a$.
From here, we can also calculate the second derivative of $E$ as 
\ba
E''(\tau,\vk)
=\:&
\frac{1}{k^2}
\frac{\partial \left[aHf\delta_m(\tau,\vk)\right]}{\partial\tau}
\nonumber
\\
=\:&
\frac{1}{k^2}a^2H^2\left[\frac{3}{2}\Omega_m-f\right]\delta_m(\tau,\vk),
\ea
where the second equality comes from the time evolution of the 
linear density contrast (continuity and Euler equations). 
Finally, from the Einstein equation [Eq.~(21a) in Ref.~\cite{ma/bertschinger:1995}],
we calculate $\phi(\tau,\vk)=-\eta(\tau,\vk)$ as
\begin{align}
\phi (\tau,\vk)
=\:&
-\frac{1}{2k^2}
\left[
aH h' (\tau,\vk)
+ 8\pi Ga^2\delta T^0_0
\right]
\nonumber
\\
=\:&
\nonumber
( a^2H^2f + 4\pi G \bar{\rho}_m a^2 ) \frac{\delta_m(\tau,\vk)}{k^2}
\\
=\:&
a^2H^2
\left( f +\frac{3}{2} \Omega_m \right)
\frac{\delta_m(\tau,\vk)}{k^2}.
\end{align}
Note that while $D$ contains a spatial gauge mode, 
$\phi\equiv D-\nabla^2E/3$ does not contain any.
Also, from \refeq{etaprime}, it is obvious that $\phi$ is constant
in time. The physical interpretation is that for a plane wave perturbation,
two neighboring test particles separated by an infinitesimal distance perpendicular to $\vk$
have a separation that is proportional to the background $a(t)$; only the component of separation parallel to $\vk$ is perturbed.

%%%%%%%%%%%%%%%%%%%%%%%%%%%%%%%%%%%%%%%%%%%%%%%%%%%%%%%%%%%%%%%%%%%%%%%%%%%%%
\section{Derivation of \refeq{dgreal}}
\label{app:Pgderiv}

This section outlines the derivation of \refeq{dgreal} from the
expression for $\tilde\d_g$ [\refeq{dgsc}].  We begin by deriving 
a compact expression for $\d z$, before moving on to the convergence $\hat\k$
and the derivation of \refeq{dgreal}.  

% % % % % %  % % % % % %% % % % % % % % % % % % % % % % % % %% % % % % % %% 
\subsection{Integrated Sachs-Wolfe term in synchronous comoving gauge}
\label{app:ISW}

Here we identify the ISW term in synchronous-comoving gauge
and derive a simplified expression for $\d z$.  
In conformal Newtonian gauge, the ISW term is given by
\be
\d z_{\rm ISW} =
-\int_0^{\tilde\chi}
d\chi (\Phi+\Psi)',
\ee
where 
$\Psi$ and $\Phi$ are the Bardeen potentials and prime denotes a derivative
with respect to the conformal time $\tau$.
In synchronous comoving gauge, they are
\ba
\Psi =& -aHE'-E'' {\rm~~and}\nonumber \\
\Phi =& - D + \frac{1}{3}\nabla^2 E + aHE',
\label{eq:bardeenPsiPhi}
\ea
and the ISW term becomes
\be
\d z_{\rm ISW} =
\int_0^{\tilde\chi}
d\chi \left( D'-\frac{1}{3}\nabla^2E' + E'''\right) .
\ee
Using \refeq{deltaz} and the definition of $E$, we find that the 
redshift perturbation in the same gauge is given by
\begin{equation}
\delta z 
= 
\int_0^{\tilde\chi}
d\chi
\left(
D-\frac{1}{3}\nabla^2 E + \partial_\parallel^2 E
\right)'.
\end{equation}
Now, we use that $\partial_\chi = \partial_\parallel - \partial_\tau$, 
and rewrite the redshift perturbation by successively applying 
integration by parts:
\begin{align}
\delta z 
=\:&
\int_0^{\tilde\chi}
d\chi
\left[
D'-\frac{1}{3}\nabla^2 E'
+ 
\left( \partial_\chi + \partial_\tau \right)
\partial_\parallel E'
\right]
\nonumber
\\
=\:&
\left[\partial_\parallel E'\right]^s_o
+
\int_0^{\tilde\chi}
d\chi
\left[
D'-\frac{1}{3}\nabla^2 E'
+ 
\left( \partial_\chi + \partial_\tau \right)
E''
\right]
\nonumber
\\
=\:&
\left[
\partial_\parallel E'
+
E''\right]^s_o
+
\int_0^{\tilde\chi}
d\chi 
\left(
D' - \frac{1}{3}\nabla^2 E'
+ 
E'''
\right).
\nonumber\\&
\label{eq:deltaz_sc}
\end{align}
The second, integral term is clearly equal to the ISW contribution, 
and neglecting it we obtain $\delta z=[\partial_\parallel E'+E'']^s_o$,
which in Fourier space becomes
\be
\delta z = ik\mu E' + E''.
\ee
Here, we have ignored the contribution from the origin, and 
$\mu$ denotes the directional cosine
between the wave vector and the line of sight direction.

% % % % % %  % % % % % %% % % % % % % % % % % % % % % % % % %% % % % % % %% 
\subsection{Convergence in synchronous comoving gauge}
\label{app:kappa}

In this section, we calculate the coordinate convergence in the 
synchronous-comoving gauge defined in \refeq{khat}. 
By applying the perpendicular derivative $\partial_{\perp i}$ to the
perpendicular directional displacement $\Delta x_\perp^i$ in \refeq{Dxperp2},
we find 
\ba
\hat\k  =\:& 
 -\frac{1}{2} \nabla_\perp^2 
\int_0^{\chit} d\chi (\chit-\chi)\frac{\chit}{\chi}
\left(D + E_\parallel \right)\vs
 & + \partial_{\perp\,i} \int_0^{\chit} d\chi \frac{\chit}{\chi}
     \left(E_j^i\nhatt^j - E_\parallel\nhatt^i\right)
\vs &
 -\frac12\partial_{\perp i}\left\{ \tilde\chi[ E_{ij}(o)\nhatt^j - E_{jk}(o)\nhatt^i\nhatt^j\nhatt^k]\right\}
\nonumber \\
\equiv\:& \hat\k^{(1)} +\hat\k^{(2)}+\hat\k^{(3)},
\label{eq:khat3}
\ea
where we have labelled the three terms and used ``$(o)$'' to denote metric shear evaluated at the observer.
In the first term, we have pulled out the perpendicular derivative
inside the integral over the unperturbed geodesic.  Hence the additional 
factor of $\chit/\chi$ in the integrand.  

We begin with the $\hat\k^{(1)}$ term:
\ba
\hat\k^{(1)}
=\:&
-\frac{1}{2} \nabla_\perp^2 \int_0^{\chit} d\chi (\chit-\chi)\frac{\chit}{\chi}
\left(D - \frac{1}{3}\nabla^2 E + \partial_\parallel^2 E \right)\vs
=\:& \kappa - \frac{1}{2} \nabla_\perp^2 \int_0^{\chit} d\chi (\chit-\chi)\frac{\chit}{\chi}
(\partial_\parallel^2 E - E'').
\ea
This allows us to relate $\hat\k$ to $\kappa$ with the introduction of some new terms. We can combine these terms with terms in $\hat\k^{(2)}$ if we simplify the latter: using the commutation relations \refeqs{comm1}{commn}, we have
\ba
\nhatt^j E_j^i- &\nhatt^i E_\parallel\vs
=\:&
\nhatt^j 
\left(
\partial_j\partial^i
-\frac{1}{3}\delta^i_j\nabla^2
\right)E
- 
\nhatt^i
\left(
\partial_\parallel^2
-\frac{1}{3}\nabla^2
\right)E\vs
=\:&
\left (\nhatt^j \partial_j\partial^i - \nhatt^i \partial_\parallel^2 \right) E
= \left( \partial_\parallel\partial^i - \partial_\parallel \nhatt^i 
\partial_\parallel \right )E\vs
=\:& \partial_\parallel \partial_\perp^i E
= \partial_\perp^i\partial_\parallel E - \frac1\chi\partial_\perp^iE,
\ea
and so
\be
\hat\k^{(2)}
= \nabla^2_{\perp} \int_0^{\chit} d\chi 
\left( \frac{\chit^2}{\chi^2} \partial_\parallel E - \frac{\chit^2}{\chi^3} E \right).
\ee
(Note again the additional factor of $\chit/\chi$ that arises since when we move $\partial_\perp^i$ outside the integral it acts at radius $\tilde\chi$ rather than $\chi$.)  Combining these gives
\begin{eqnarray}
\hat\k^{(1)}+\hat\k^{(2)} &=& \kappa + \frac12\nabla^2_\perp \int_0^{\chit} d\chi 
\Bigl[
\Bigl(-\frac{\chit^2}\chi+\chit\Bigr)
(\partial_\parallel^2 E - E'')
\nonumber \\
&& + 2\frac{\chit^2}{\chi^2} \partial_\parallel E - 2\frac{\chit^2}{\chi^3} E \Bigr]
\nonumber \\
&=& \kappa + \frac12\nabla^2_\perp \int_0^{\chit} d\chi \nonumber \\ && \times
\Bigl[
\Bigl(-\frac{\chit^2}\chi+\chit\Bigr)\frac{d}{d\chi} 
\left(2\partial_\parallel E - \frac{dE}{d\chi}\right)
\nonumber \\
&& \quad + 2\frac{\chit^2}{\chi^2} \partial_\parallel E - 2\frac{\chit^2}{\chi^3} E \Bigr],
\end{eqnarray}
where in the second line we have used $' = \partial_\parallel - d/d\chi$. The $\partial_\parallel E$ terms form a total derivative, which may be separately evaluated:
\begin{eqnarray}
\hat\k^{(1)}+\hat\k^{(2)} &=& \kappa
+ \frac12\nabla^2_\perp \int_\epsilon^{\chit} d\chi 
\Bigl[
\Bigl(\frac{\chit^2}\chi-\chit\Bigr) \frac{d^2E}{d\chi^2}
- 2\frac{\chit^2}{\chi^3} E \Bigr]
\nonumber \\
&& + \nabla_\perp^2\left\{\left. \Bigl(-\frac{\chit^2}\chi+\chit\Bigr)\partial_\parallel E \right|_\epsilon^{\chit} \right\}.
\end{eqnarray}
The quantity in braces vanishes at $\chit$, whereas at $\epsilon\rightarrow 0$ it blows up. For this reason, we will evaluate this expression only for $\epsilon>0$ and then take the limit after all divergences are cancelled. If we Taylor expand $\partial_\parallel E$ to order $\epsilon$, we find
\be
\partial_\parallel E(\epsilon) =
\nhatt^i E_{,i}(o) + \nhatt^i\nhatt^j\epsilon E_{,ij}(o) - \nhatt^i E'_{,i}(o)\epsilon
+{\cal O}(\epsilon^2).
\ee
The quantity in braces is then (keeping terms that are nonvanishing as $\epsilon\rightarrow 0^+$)
\be
-\left(-\frac{\tilde\chi^2}\epsilon + \tilde\chi \right) E_{,i}(o)\nhatt^i
+ \tilde\chi^2 \nhatt^i\nhatt^j E_{,ij}(o)
- \tilde\chi^2 \nhatt^i E'_{,i}(o).
\ee
The $\nabla^2_\perp$ operator pulls down a factor of $0$ for monopoles, $-2/\tilde\chi^2$ for dipoles, and $-6/\tilde\chi^2$ for quadrupoles. It follows that
\begin{eqnarray}
\hat\k^{(1)}+\hat\k^{(2)} &=& \kappa
+ \frac12\nabla^2_\perp \int_\epsilon^{\chit} d\chi 
\Bigl[
\Bigl(\frac{\chit^2}\chi-\chit\Bigr) \frac{d^2E}{d\chi^2}
- 2\frac{\chit^2}{\chi^3} E \Bigr]
\nonumber \\
&& + \left(-\frac2\epsilon + \frac2{\chit}\right)E_{,i}(o)\nhatt^i - 6E_{ij}(o)\nhatt^i\nhatt^j 
\nonumber \\ && + 2\nhatt^i E'_{,i}(o).
\end{eqnarray}
The remaining integral is also a total derivative, although in this case two integrations by parts are necessary. Both terms can be integrated by parts to obtain a single derivative $dE/d\chi$; the integrals cancel leaving only the boundary terms:
\begin{eqnarray}
\hat\k^{(1)}+\hat\k^{(2)} &=& \kappa
+ \frac12\nabla^2_\perp \Bigl[
\left. \Bigl(\frac{\chit^2}\chi-\chit\Bigr)\frac{dE}{d\chi}\right|_\epsilon^{\chit}
+ \left.\frac{\tilde\chi^2}{\chi^2}E\right|_\epsilon^{\chit} \Bigr]
\nonumber \\
&& + \left(-\frac2\epsilon + \frac2{\chit}\right)E_{,i}(o)\nhatt^i - 6E_{ij}(o)\nhatt^i\nhatt^j 
\nonumber \\ && + 2\nhatt^i E'_{,i}(o).
\end{eqnarray}
This simplifies to
\begin{eqnarray}
\hat\k^{(1)}+\hat\k^{(2)} &=& \kappa + \frac12\nabla_\perp^2 \Bigl[ -\left(\frac{\tilde\chi^2}\epsilon - \tilde\chi\right)\frac{dE}{d\chi}(\epsilon)
- \frac{\tilde\chi^2}{\epsilon^2}E(\epsilon) \Bigr]
\nonumber \\
&& + \frac12\nabla^2_\perp E
 + \left(-\frac2\epsilon + \frac2{\chit}\right)E_{,i}(o)\nhatt^i 
\nonumber \\ && - 6E_{ij}(o)\nhatt^i\nhatt^j 
+ 2\nhatt^i E'_{,i}(o),
\label{eq:temp-0325}
\end{eqnarray}
where $\nabla_\perp^2$ is evaluated at radius $\chit$.

Further simplification requires the limiting forms of the terms in Eq.~(\ref{eq:temp-0325}). A lowest-order expansion gives the quantity in brackets as
\begin{eqnarray}
&& -\left(\frac{\tilde\chi^2}\epsilon-\tilde\chi\right)[\nhatt^i E_i(o) - E'(o) + \epsilon \nhatt^i\nhatt^j E_{ij}(o)
\nonumber \\ && \;\;\;\; - 2\epsilon \nhatt^iE'_{,i}(o) + \epsilon E''(o)]
\nonumber \\ &&
-\frac{\tilde\chi^2}{\epsilon^2}[E(o) + \epsilon \nhatt^iE_{,i}(o) - \epsilon E'(o)
\nonumber \\ && \;\;\;\;
+ \frac12\epsilon^2\nhatt^i\nhatt^j E_{,ij}(o) - \epsilon^2\nhatt^i E'_{,i}(o) + \frac12\epsilon^2 E''(o)
].~~~~~~
\end{eqnarray}
Again using that
$\nabla^2_\perp$ operator pulls down a factor of $0$ for monopoles, $-2/\tilde\chi^2$ for dipoles, and $-6/\tilde\chi^2$ for quadrupoles, we find that Eq.~(\ref{eq:temp-0325}) simplifies to
\begin{eqnarray}
\hat\k^{(1)}+\hat\k^{(2)} &=& \kappa
+ \frac12\nabla^2_\perp E
 + \frac1{\chit}E_{,i}(o)\nhatt^i 
\nonumber \\ && - \frac32 E_{ij}(o)\nhatt^i\nhatt^j 
-\nhatt^i E'_{,i}(o).
\label{eq:temp-0344}
\end{eqnarray}

The third term, $\hat\k^{(3)}$, is independent of $\tilde\chi$ since it is a derivative of a quantity that depends linearly on $\tilde\chi$; hence we may evaluate it on the unit sphere $\tilde\chi=1$. We find
\begin{eqnarray}
\hat\k^{(3)} &=& -\frac12E_{ij}(o)(\delta_{ij}-\nhatt^i\nhatt^j) + \frac12E_{jk}(o)\bigl[2\nhatt^j\nhatt^k
\nonumber \\ &&
+ (\delta_{ij}-\nhatt^i\nhatt^j)\nhatt^i\nhatt^k
+ (\delta_{ik}-\nhatt^i\nhatt^k)\nhatt^i\nhatt^j\bigr]
\nonumber \\
&=& \frac32 E_{ij}(o)\nhatt^i\nhatt^j.
\end{eqnarray}
Combining with Eq.~(\ref{eq:temp-0344}) gives
\begin{equation}
\hat\k = \kappa
+ \frac12\nabla^2_\perp E
 + \frac1{\chit}E_{,i}(o)\nhatt^i 
-\nhatt^i E'_{,i}(o).
\label{eq:hatk-1}
\end{equation}
This is \refeq{khat2}.

% % % % % %  % % % % % %% % % % % % % % % % % % % % % % % % %% % % % % % %% 
\subsection{From \refeq{nobssc} to \refeq{dgsc} and \refeq{dgreal}}
\label{app:derivdg}

Let us start from \refeq{nobssc},
\ba
\nonumber
\tilde{\delta}_g(\tilde\vx)
=\:& \d_g + \bt\d z
+ 2 D - E_\parallel \vs
& + 2\frac{\D x_\parallel}{\chit} - 2\hat\k 
- \left(1 - \frac{1+\zt}{H}\frac{d H(\zt)}{d\zt}
\right) \d z \vs
& - \frac{1+\zt}{H(\zt)} (D' + E'_\parallel)\Big|_{\chit},
\label{eq:dgobs_sc}
\ea
where 
\be
\Delta x_\parallel
=
-\int_0^{\tilde\chi}d\chi
(D+E_\parallel)
-\frac{1+\tilde{z}}{H(\tilde{z})}\delta z
\ee
and we use \refeq{hatk-1} for $\hat\k$.
Also, in the previous section, 
we found that the redshift perturbation, ignoring the ISW term, is given by
\be
\delta z = \partial_\parallel E' + E''.
\label{eq:delta-z.0}
\ee

As shown in \refapp{sc2cN}, $D$ and $E$ contain spatial gauge modes,
while $\phi \equiv D-\nabla^2 E/3$ and $E'$ remove such gauge modes.
We now collect the four terms in \refeq{dgobs_sc} which contain gauge modes:
\ba
&2D-E_{\parallel}+2\frac{\Delta x_\parallel}{\tilde\chi}-2\hat\kappa
\nonumber
\\
&\;\;=
2\phi 
+\frac{2}{3}\nabla^2E
-
\left(
\partial^2_\parallel -\frac{1}{3}\nabla^2
\right)E
-
\frac{2}{\chit}
\int_0^{\tilde\chi}d\chi
(D+E_\parallel)
\vs
&\;\;\;\;\;\;-\frac{2}{\chit}\frac{1+\tilde{z}}{H(\tilde{z})}\delta z
-2\k-\nabla_{\perp}^2E 
- \frac2{\chit}\nhatt^iE_{,i}(o) + 2\nhatt E'_{,i}(o)
\vs
&\;\;=
2\phi 
+
\frac{2}{\chit}[\partial_\parallel E - \partial_\parallel E(o)]
-
\frac{2}{\chit}
\int_0^{\tilde\chi}d\chi
(\phi + \partial_\parallel^2 E)
\vs
&\;\;\;\;\;\;-\frac{2}{\chit}\frac{1+\tilde{z}}{H(\tilde{z})}\delta z -2\k + 2\partial_\parallel E'(o).
\ea
The third term can be further simplified by double integration by parts to yield
\begin{eqnarray}
\int_0^{\tilde\chi}d\chi (\phi + \partial_\parallel^2 E )
&=&
\int_0^{\tilde\chi}d\chi (\phi +  E'') + E' + \partial_\parallel E
\nonumber \\ &&
- E'(o) - \partial_\parallel E(o),
\label{eq:integ-dip}
\end{eqnarray}
Then we find for the observed galaxy density contrast
\ba
\nonumber
\tilde{\delta}_g(\tilde\vx)
=\:& \d_g + \bt\d z
- \frac{1+\zt}{H(\zt)} \partial_\parallel^2E'
\vs
&- \left[
1 - \frac{1+\zt}{H}\frac{d H(\zt)}{d\zt}
+
\frac{2}{\chit}\frac{1+\tilde{z}}{H(\tilde{z})}
\right] \d z +2\phi 
\vs
&
-\frac{2}{\chit}[E'-E'(o)]
-\frac{1+\tilde{z}}{H(\tilde{z})}\phi'
-\frac{2}{\chit}\int_0^{\tilde\chi}d\chi\left(\phi + E''\right) 
\vs &
- 2\k + 2\partial_\parallel E'(o).
\ea
This is \refeq{dgsc}.  Note that the magnification bias contribution $\Q\, \d \M$
does not contain gauge modes [\refeq{magsc}].  

To proceed to \refeq{dgreal}, we drop several terms:
\begin{list}{$\bullet$}{}
\item
The observer terms (which contribute only to the monopole and dipole).
\item
The $\phi'$ term, since the $0i$ component of the Einstein equation ensures that in a $\Lambda$CDM universe $\phi'=0$ 
[see \refeq{etaprime} and note that $\eta =-\phi$].
\item
The terms $-\frac{2}{\chit}\int_0^{\tilde\chi}d\chi(\phi + E'')$ in \refeq{dgsc} and
\refeq{magsc}, corresponding to the time delay which is very 
small \cite{HuCooray00,DodelsonEtal08}.  
\item
The terms involving the convergence $\k$, which is generally {\em not} small but is a projected quantity, and in the flat-sky limit contributes only to transverse modes.
\end{list}

Substituting in \refeq{delta-z.0} for $\delta z$ and using \refeq{magsc} then yields \refeq{dgreal}.  Conversion to Fourier space in the flat-sky limit -- i.e. where we make the replacement $\partial_\parallel \rightarrow ik\mu$ -- then gives \refeq{dgFourier}.

%%%%%%%%%%%%%%%%%%%%%%%%%%%%%%%%%%%%%%%%%%%%%%%%%%%%%%%%%%%%%%%%%%%%%%%%%%%%%
\section{Test cases for the observed galaxy overdensity}
\label{app:test}

This appendix considers several analytical test cases that serve as a cross-check
of \refeq{dgsc}.  The first two cases are pure gauge modes, with the
expected result that $\tilde\d_g$ does not receive any contributions from
such perturbations.  We then consider a perturbed expansion history, spatial
curvature, and a Bianchi type~I cosmology with anisotropic expansion. Finally we consider a model with a time-dependent linear gradient in $E$, which has no metric perturbation but leads to a delicate cancellation of terms in the galaxy density.  In
all cases, the linearized version of the exact result can be derived straightforwardly,
and we show that it agrees with the prediction of \refeq{dgsc} in all
cases. We further evaluate the magnification $\mmu$ using \refeq{magsc} and show that it matches the expected results.

\subsection{Pure spatial gauge mode}

The residual gauge freedom of the synchronous gauge allows us to reparameterize the spatial coordinates according to $x^i\rightarrow x^i+\xi^i$, where $\xi^i$ depends only on the spatial coordinates and not $x^0$. This leads to a spatial metric perturbation $h_{ij} = -a^2(\xi^i{_{,j}}+\xi^j{_{,i}})$. Since here we consider scalar perturbations only, $\xi^i$ should be derived from a potential $\xi^i=\xi_{,i}$. If we start from an unperturbed Universe, the resulting metric perturbation is
\begin{equation}
D = -\frac13\nabla^2\xi {\rm ~~and~~} E = -\xi.
\label{eq:PG}
\end{equation}
Eq.~(\ref{eq:PG}) corresponds to a pure gauge mode.

For this mode, we proceed to evaluate $\delta z$, $\phi$, and $\kappa$. Since $D$ and $E$ are time-independent, it is trivially seen that $\delta z=0$, and Eq.~(\ref{eq:PG}) immediately implies $\phi=0$. Finally, since $D=\frac13\nabla^2E$ and $E''=0$ we can also see that $\kappa=0$. Then Eq.~(\ref{eq:dgsc}) reduces to
\begin{equation}
\tilde\delta_g(\tilde{\bf x}) = \delta_g,
\end{equation}
which is the expected answer. That is, in this case we do not have any contributions to the galaxy density aside from the intrinsic contribution.

We may also evaluate the magnification using \refeq{magsc}. With $E'=0$ everywhere and $\delta z = \phi = \kappa = 0$, it is trivially seen that $\delta\mmu=0$.

\subsection{Zero-wavenumber gauge mode}

There is another spatial gauge mode that does not fall into the rubric of Eq.~(\ref{eq:PG}): the zero-wavenumber mode given by
\begin{equation}
D = \Xi + g_ix^i {\rm ~~and~~} E=0,
\label{eq:PZ}
\end{equation}
where $\Xi$ is a constant scalar and ${\bf g}$ is a constant vector field. This is generated by the gauge perturbation
\begin{equation}
\xi^i = -\Xi x^i + \frac12g^i |{\bf x}|^2 - g_jx^jx^i.
\end{equation}
Again we trivially have $\delta z=0$, but this time $\phi=\Xi+\chi{\bf g}\cdot\hat{\bf n}$. Also we have
\begin{equation}
\nabla_\perp^2D = \nabla_\perp^2(\chi{\bf g}\cdot\hat{\bf n}) = \frac{-2}\chi{\bf g}\cdot\hat{\bf n},
\end{equation}
since ${\bf g}\cdot\hat{\bf n}$ is a dipole ($\ell=1$) and for a pure multipole of order $\ell$ the operator $\nabla_\perp^2$ yields a factor of $-\ell(\ell+1)/\chi^2$. Thus we find
\begin{equation}
\kappa = -\frac12\int_0^{\tilde\chi} d\chi (\tilde\chi-\chi)\frac\chi{\tilde\chi}\frac{-2}\chi{\bf g}\cdot\hat{\bf n}
= \frac12 \tilde\chi {\bf g}\cdot\hat{\bf n}.
\end{equation}

The galaxy density perturbation obtained via Eq.~(\ref{eq:dgsc}) has only four nontrivial terms:
\begin{equation}
\tilde\delta_g = \delta_g + 2(\Xi+\tilde\chi{\bf g}\cdot\hat{\bf n}) - \frac2{\tilde\chi} \int_0^{\tilde\chi} (\Xi+\chi{\bf g}\cdot\hat{\bf n})\,d\chi
- \tilde\chi {\bf g}\cdot\hat{\bf n}.
\end{equation}
Here the second term comes from the $+2\phi$ term in Eq.~(\ref{eq:dgsc}), the third term is the line of sight integral of $\phi+E''=\phi$, and the last term comes from the $-2\kappa$. It is easily seen that these three terms cancel, leaving $\tilde\delta_g = \delta_g$, which is the expected answer.

Unlike the previous case, here the magnification contains nontrivial terms: substituting the nonzero values of $\phi$ and $\kappa$ into \refeq{magsc}, we find
\be
\d\mmu = -2(\Xi+\tilde\chi{\bf g}\cdot\hat{\bf n}) + \tilde\chi{\bf g}\cdot\hat{\bf n} + \frac2{\tilde\chi} \int_0^{\tilde\chi} (\Xi+\chi{\bf g}\cdot\hat{\bf n}) \,d\chi
=0,
\ee
as expected.

\subsection{Perturbation to the expansion history}

A less trivial type of perturbation is one in which we alter the cosmic expansion rate. This can be done by setting
\begin{equation}
D = D(\tau) {\rm ~~and~~} E=0.
\label{eq:PE}
\end{equation}
The Universe so described is still an FRW model since it is homogeneous and isotropic. (It may no longer be a solution to the Friedmann equation with only matter+$\Lambda$, however this does not concern us since we are testing an equation derived only using kinematics.) However, it has a ``true'' scale factor $a_{\rm true}$ that is related to the unperturbed scale factor via
\begin{equation}
a_{\rm true}(\tau) = a(\tau) [1+D(\tau) - D(\tau_0)],
\label{eq:a-true}
\end{equation}
where we fix $a_{\rm true}$ to be unity today. The true time coordinate (proper time in the case of FRW) remains equal to the coordinate time,
\begin{equation}
t_{\rm true} = t = \int a(\tau)\,d\tau.
\end{equation}
The true conformal time is then
\begin{equation}
\tau_{\rm true} = \int \frac{dt_{\rm true}}{a_{\rm true}} = \int \frac{a\,d\tau}{a_{\rm true}}
= \int [1-D(\tau)+D(\tau_0)]\,d\tau.
\end{equation}
(The integration constant is chosen to set $\tau_{\rm true}=0$ at the Big Bang, but we do not need to make use of this fact.) Integrating gives
\begin{equation}
\tau_{\rm true,0}-\tau_{\rm true} = [1+D(\tau_0)](\tau_0-\tau) - \int_{\tau}^{\tau_0} D(\tau_1)\,d\tau_1.
\label{eq:diff-tau}
\end{equation}

We care in particular about the behavior as a function of the observed redshift $z$, which is related to $a_{\rm true}=(1+\tilde z)^{-1}$. It follows from Eq.~(\ref{eq:a-true}) that the comoving distance relation is now
\begin{equation}
a(\tau) = a_{\rm true}(\tau) [1-D(\tau)+D(\tau_0)]
\end{equation}
and so we may write the perturbation to the conformal time,
\begin{equation}
\tau(a_{\rm true}) = \tau_{\rm bg}(a_{\rm true}) + \frac{D(\tau_0)-D(\tau)}{a_{\rm true}H(a_{\rm true})}.
\end{equation}
Here $\tau_{\rm bg}$ is the background conformal time-scale factor relation.  Finally 
using Eq.~(\ref{eq:diff-tau}) yields
\begin{eqnarray}
\tau_{\rm true,0}-\tau_{\rm true}
 &=&  [1+D(\tau_0)] [ \tau_{\rm bg}(1)-\tau_{\rm bg}(a_{\rm true})]
\nonumber \\ &&  -  \frac{D(\tau_0)-D(\tau)}{a_{\rm true}H(a_{\rm true})}
- \int_{\tau}^{\tau_0} D(\tau_1)\,d\tau_1.
\nonumber \\ &&
\end{eqnarray}
This is the true comoving radial distance $\chi_{\rm true}$ to redshift $a_{\rm true}^{-1}-1$. That is,
\begin{eqnarray}
\chi_{\rm true} &=& [1+D(\tau_0)] \chi_{\rm bg}(a_{\rm true})
\nonumber \\ &&  -  \frac{D(\tau_0)-D(\tau)}{a_{\rm true}H(a_{\rm true})}
- \int_{\tau}^{\tau_0} D(\tau_1)\,d\tau_1.
\label{eq:chi-true}
\end{eqnarray}
The true Hubble rate at this time is
\begin{eqnarray}
H_{\rm true} &=& \frac{d\ln a_{\rm true}}{dt}
\nonumber \\
&=& H(\tau) + \frac{D'(\tau)}a
\nonumber \\
&=& H_{\rm bg}(a_{\rm true}) + \frac{dH}{d\tau} [ \tau(a_{\rm true})-\tau_{\rm bg}(a_{\rm true}) ] + \frac{D'(\tau)}a
\nonumber \\
&=& H_{\rm bg}(a_{\rm true}) + \frac{dH}{d\tau} \frac{D(\tau_0)-D(\tau)}{a_{\rm true}H(a_{\rm true})} + \frac{D'(\tau)}a
\nonumber \\
&=& H_{\rm bg}(a_{\rm true}) + a\frac{dH}{da} [D(\tau_0)-D(\tau)] + \frac{D'(\tau)}a.
\nonumber \\ &&
\end{eqnarray}

We expect the perturbation in the observed galaxy density to have several parts: there is a perturbation in the physical galaxy density, a part associated with the different epoch in cosmic history at which the galaxy density is measured (different $t$; one wants the different physical density here so we include both the $b_e$ evolution and the $-3$ associated with the dilution of comoving volume), and a part associated with the different physical volume. Specifically:
\begin{equation}
\tilde \delta_g = \delta_g
  + (b_e-3) [D(\tau_0)-D(\tau)]
+ \ln\frac{dV_{\rm true}/da_{\rm true}\,d\Omega}{dV/da\,d\Omega}.
\end{equation}
The comoving volume effect is computable from Eq.~(\ref{eq:chi-true}). We see that
\begin{eqnarray}
\frac{dV_{\rm true}}{da_{\rm true}\,d\Omega} &=& \frac{a_{\rm true}^2\chi^2_{\rm true}}{H_{\rm true}}
\nonumber \\
&=& \frac{a_{\rm true}^2[\chi_{\rm bg}(a_{\rm true})]^2}{H_{\rm bg}(a_{\rm true})} \Bigl\{ 1+2D(\tau_0)
\nonumber \\ &&
- \frac{2\int_{\tau}^{\tau_0} D(\tau_1)d\tau_1}{\chi(a_{\rm true})}
- \frac{2[D(\tau_0)-D(\tau)]}{a_{\rm true}H(a_{\rm true})\chi(a_{\rm true})}
\Bigr\}
\nonumber \\ &&
\times\Bigl\{ 1 - \frac aH\frac{dH}{da}[D(\tau_0)-D(\tau)]  - \frac{D'(\tau)}{aH} \Bigr\}.
\nonumber \\ &&
\end{eqnarray}
This leads to
\begin{eqnarray}
\tilde \delta_g &=& \delta_g + (b_e-3) [D(\tau_0)-D(\tau)]
 + 2D(\tau_0)
 \nonumber \\ &&
 + \left( - \frac{d\ln H}{d\ln a} - \frac2{aH\chi} \right) [D(\tau_0)-D(\tau)]
\nonumber \\ &&
- \frac{2\int_{\tau}^{\tau_0} D(\tau_1)d\tau_1}{\chi(a_{\rm true})} - \frac{D'(\tau)}{aH}.
\label{eq:result-2}
\end{eqnarray}
The expected ``magnification'' $\delta\mmu$ is twice the perturbation to the angular diameter distance at fixed observed redshift; in a flat universe this is equivalent to the perturbation to $\chi_{\rm true}$. Using Eq.~(\ref{eq:chi-true}),
\begin{eqnarray}
\d\mmu &=& -2\frac{\chi_{\rm true}-\chi_{\rm bg}}{\chi_{\rm bg}}
\nonumber \\
&=& -2D(\tau_0) + \frac{2[D(\tau_0)-D(\tau)]}{aH\chi_{\rm bg}}
\nonumber \\ &&
 + \frac2{\chi_{\rm bg}}\int_\tau^{\tau_0} D(\tau_1)d\tau_1.
\label{eq:expect-2}
\end{eqnarray}

In comparison, if we use Eq.~(\ref{eq:dgsc}) to find $\tilde\delta_g$, then we find that the perturbation in Eq.~(\ref{eq:PE}) yields $\phi=D(\tau)$, $\kappa=0$, and $\delta z = D(\tau_0)-D(\tau)$. Therefore,
\begin{eqnarray}
\tilde\delta_g &=& \delta_g + \left( b_e - 1 - \frac{d\ln H}{d\ln a} - \frac2{aH\chi} \right) [D(\tau_0)-D(\tau)]
\nonumber \\ &&
+ 2D(\tau) - \frac{2\int_{\tau}^{\tau_0} D(\tau_1) d\tau_1}{\chi}
-\frac{D'(\tau)}{aH}.
\end{eqnarray}
A simple comparison shows this to be equivalent to Eq.~(\ref{eq:result-2}). Similarly, evaluation of \refeq{magsc} gives
\begin{eqnarray}
\d\mmu &=& -2D(\tau) + \frac2{\tilde\chi}\int_\tau^{\tau_0} D(\tau_1)d\tau_1
\nonumber \\ &&
 + \left( -2 + \frac2{aH\chi}\right)[D(\tau_0)-D(\tau)],
\end{eqnarray}
in agreement with Eq.~(\ref{eq:expect-2}).

\subsection{Spatial curvature}

A fourth example of a perturbation we consider is spatial curvature. Under stereographic projection, a 3-sphere of curvature $K$ (radius of curvature $K^{-1/2}$) can be written with 3-metric
\begin{equation}
ds_3^2 =  \left(1 + \frac14K |{\bf x}|^2 \right)^{-2} dx^i dx^i,
\end{equation}
or to first order in $K$,
\begin{equation}
D = -\frac14K|{\bf x}|^2 {\rm ~~and~~} E=0.
\label{eq:PK}
\end{equation}
The expected result is that in the perturbed universe, the radial comoving distance-redshift relation remains the same. However, there is a change in the volume element associated with the change in the comoving angular diameter distance,
\begin{equation}
\frac{dV_{\rm new}}{dV_{\rm old}} = \frac{\sin_K^2\tilde\chi}{\tilde\chi^2} = 1-\frac13K\tilde\chi^2 + {\cal O}(K^2),
\end{equation}
where $\sin_K$ is the sinelike function:
\begin{equation}
\sin_K \chi = \left\{\begin{array}{lll}
\chi & & K=0 \\
K^{-1/2} \sin (K^{1/2}\chi) & & K>0 \\
(-K)^{-1/2} \sinh [(-K)^{1/2}\chi] & & K<0.
\end{array}\right.
\end{equation}
Thus we expect to obtain
\begin{equation}
\tilde\delta_g = \delta_g - \frac13K\tilde\chi^2.
\label{eq:expect-PK}
\end{equation}
The magnification $\d\mmu$ is $-2$ times the fractional perturbation to the angular diameter distance coming from spatial curvature, which is
\be
\d\mmu = -2\frac{\sin_K\tilde\chi - \tilde\chi}{\tilde\chi} = \frac13K\tilde\chi^2.
\label{eq:expect-PK-mag}
\ee

If we instead use Eq.~(\ref{eq:dgsc}), we find that $\phi=-\frac14K\chi^2$, $\kappa=0$ (since $D$ is a pure monopole, $\nabla_\perp^2D=0$ even though $\nabla^2D\neq0$), and $\delta z=0$. Then
\begin{equation}
\tilde\delta_g = \delta_g -\frac12K\tilde\chi^2 -\frac2{\tilde\chi}\int_0^{\tilde\chi} \left( -\frac14K\chi^2 \right)\,d\chi.
\end{equation}
Evaluation of the integral trivially recovers Eq.~(\ref{eq:expect-PK}).

We can also compute the magnification from \refeq{magsc}; we get
\be
\d\mmu = \frac12K\tilde\chi^2 + \frac2{\tilde\chi} \int_0^{\tilde\chi} \left(-\frac14K\chi^2\right) d\chi = \frac13K\tilde\chi^2,
\ee
in agreement with \refeq{expect-PK-mag}.

\subsection{Bianchi I cosmology}

The previous test cases have not tested the terms involving $E'$. One case that does is the Bianchi I cosmology, in which the three spatial axes (usually taken to be the coordinate axes) have different scale factors but the universe is still homogeneous. We will focus here on the case where the observer looks in the $x^3$-direction and the metric perturbations are
\begin{eqnarray}
D({\bf x},\tau) &=& -s_3(\tau) {\rm~~and~~}
\nonumber \\
E({\bf x},\tau) &=& \frac{s_1(\tau)(x^1)^2 + s_2(\tau)(x^2)^2 + s_3(\tau)(x^3)^2}2, {\rm~~~~~~}
\label{eq:PB}
\end{eqnarray}
with $s_1(\tau)+s_2(\tau)+s_3(\tau)=0$ and $s_i(\tau_0)=0$. This is equivalent to a case where the expansion along the 3-axis is unperturbed, but the other two axes are perturbed: the scale factors are
\begin{eqnarray}
a_1(t) &=& a(t) [1+s_1(t)-s_3(t)], {\rm~~}
\nonumber \\
a_2(t) &=& a(t) [1+s_2(t)-s_3(t)], {\rm~~and~~}
\nonumber \\
a_3(t) &=& a(t).
\end{eqnarray}
Note that we have already considered in Eq.~(\ref{eq:PE}) the case where the global isotropic expansion $D(\tau)$ is perturbed, so no new independent tests of our results are possible by using a different function for $D$ in Eq.~(\ref{eq:PB}). Also we have considered in Eq.~(\ref{eq:PG}) the case where $E$ is a time-independent function with zero Laplacian, so there is no independent test of our result that can be obtained by allowing $s_i(t_0)\neq 0$.

It is straightforward to determine the expected change in observed galaxy density for this model. The metric components $\{g_{00},g_{03},g_{33}\}$ are not perturbed, so for the $\hat{\bf n}=(0,0,1)$ direction the past light cone is unperturbed and a given redshift $z$ corresponds to the usual distance $\chi_{\rm bg}(z)$ and $\tau_{\rm bg}(z)$. The only nontrivial effect is in the transverse dimensions and in the volume element -- the angular diameter distance is modified by the perturbed expansion rates in the 1 and 2 directions. We may determine the true angular diameter distance along the 1 axis by considering a ray projected backward from the observer with a physical angular separation $\varsigma$ from the 3-axis, i.e. in direction $\hat{\bf n}=(\varsigma,0,1)$. (We work to order $\varsigma$ so that $\sin\varsigma=\varsigma$ and $\cos\varsigma=1$.) Then the 4-momentum of such a ray with unit energy is
\begin{equation}
p_\mu = \left(-1,-\varsigma,0,-1\right).
\end{equation}
Since the metric coefficients do not depend on spatial position in this model, the spatial covariant components $p_i$ of the momentum are conserved. Then we find that the spatial position is given by
\begin{eqnarray}
x^1 &=& \int_{\tau_0}^\tau \frac{dx^1/d\lambda}{dx^0/d\lambda} d\tau_1
\nonumber \\
&=& \int_{\tau_0}^\tau \frac{[a_1(\tau_1)]^{-2}p_1}{-[a(\tau_1)]^{-2}p_0} d\tau_1
\nonumber \\
&=& -\varsigma \int_{\tau_0}^\tau [ 1 - 2s_1(\tau_1) + 2s_3(\tau_1) ] d\tau_1.
\end{eqnarray}
The physical angular diameter distance is the physical transverse distance divided by the angle subtended, i.e. $a_1(\tau)x^1/\varsigma$. That is,
\begin{eqnarray}
D_{\rm A,phys,1} &=& -a(\tau) [ 1+s_1(\tau) - s_3(\tau) ]
\nonumber \\ && \times
\int_{\tau_0}^\tau [ 1 - 2s_1(\tau_1) + 2s_3(\tau_1) ] d\tau_1
\nonumber \\
&=& a(\tau) \Bigl\{ 1 + s_1(\tau) - s_3(\tau)
\nonumber \\ &&
+ 2\frac{\int_\tau^{\tau_0} [s_3(\tau_1)-s_1(\tau_1)] d\tau_1}{\tau_0-\tau} \Bigr\}.
\end{eqnarray}
The observed galaxy overdensity then deviates from the true galaxy overdensity only by the transverse area element (since the time of observation and the longitudinal distance-redshift relation are unaffected). That is,
\begin{equation}
\tilde\delta_g = \delta_g + \ln \frac{D_{\rm A,phys,1}D_{\rm A,phys,2}}{D_{\rm A,unpert}^2}.
\end{equation}
Using that $\sum_{i=1}^3 s_i(\tau)=0$, we may simplify this to
\begin{equation}
\tilde \delta_g = \delta_g - 3s_3(\tau) + 6\frac{\int_\tau^{\tau_0} s_3(\tau_1) d\tau_1}{\tau_0-\tau}.
\label{eq:expect-PB}
\end{equation}
The magnification should also be given by the change in the transverse area element:
\be
\d\mmu = 3s_3(\tau) - 6\frac{\int_\tau^{\tau_0} s_3(\tau_1) d\tau_1}{\tau_0-\tau}.
\label{eq:expect-PB-mag}
\ee

We now wish to compare our result to Eq.~(\ref{eq:dgsc}). By construction we have on our chosen sightline $D+E_\parallel=0$, so $\delta z=0$; and since $\nabla^2E=0$ we have $\phi=-s_3$. The convergence $\kappa$ is more complicated: $D$ is spatially constant, $\nabla^2E=0$, and $E$ is a pure quadrupole ($\ell=2$) and hence $\nabla_\perp^2$ pulls down a factor of $-\ell(\ell+1)/\chi^2=-6/\chi^2$, so we find
\begin{eqnarray}
\nabla_\perp^2\left( D - \frac13\nabla^2E - E'' \right) &=& -\frac{6E''}{\chi^2} = -\frac{6(s''_3\chi^2/2)}{\chi^2}
\nonumber \\
&=& -3s''_3,
\end{eqnarray}
where the second equality is valid only on the 3-axis line of sight. Consequently the convergence is
\begin{equation}
\kappa = \frac32 \int_0^{\tilde\chi} d\chi\,(\tilde\chi - \chi) \frac{\chi}{\tilde\chi}  s''_3(\chi).
\end{equation}
Finally we have $\partial_\parallel^2E' = s_3'$, and the observer terms $E'(o)$ and $\partial_\parallel E'(o)$ both vanish. Plugging these results into Eq.~(\ref{eq:dgsc}) gives
\begin{eqnarray}
\tilde\delta_g &=& \delta_g - \frac{1+z}{H} s_3'(\tilde\chi) - 2s_3(\tilde\chi) - \frac2{\tilde\chi}\left[ \frac12\tilde\chi^2s'_3(\tilde\chi)\right]
\nonumber \\
&& - \frac2{\tilde\chi} \int_0^{\tilde\chi} d\chi \left[ -s_3(\chi) + \frac12\chi^2s''_3(\chi) \right]
\nonumber \\
&& - 3 \int_0^{\tilde\chi} d\chi\,(\tilde\chi - \chi) \frac{\chi}{\tilde\chi}  s''_3(\chi)
\nonumber \\
&& - \frac{1+z}H [-s'_3(\tilde\chi)].
\end{eqnarray}
The terms containing $s'_3(\tilde\chi)/H$ cancel and the two integrals can be combined, yielding
\begin{eqnarray}
\tilde\delta_g &=& \delta_g - 2s_3(\tilde\chi) - \tilde\chi s'_3(\tilde\chi)
\nonumber \\ &&
+ \int_0^{\tilde\chi} d\chi \left[ \frac2{\tilde\chi}s_3(\chi) + \left(2\frac{\chi^2}{\tilde\chi}-3\chi\right) s''_3(\chi) \right].~~~~
\label{eq:temp-e5}
\end{eqnarray}
This does not quite resemble Eq.~(\ref{eq:expect-PB}), but we can cast it in a similar form by applying repeated integration by parts to the second derivative term. For a general function $f$,
\begin{eqnarray}
\int_0^{\tilde\chi} f(\chi) s''_3(\chi) d\chi &=& -f(\tilde\chi) s'_3(\tilde\chi) + f(0)s'_3(0)
\nonumber \\ &&
+ f'(\tilde\chi)s_3(\tilde\chi) - f'(0)s_3(0)
\nonumber \\ &&
+ \int_0^{\tilde\chi} f''(\chi) s_3(\chi) d\chi,
\end{eqnarray}
where the unusual signs result from the fact that $'$ denotes a derivative with respect to $\tau$ instead of $\chi$ (the relation is simply a minus sign). Then, recalling that $s_3(\chi=0)=0$, Eq.~(\ref{eq:temp-e5}) simplifies to
\begin{equation}
\tilde\delta_g = \delta_g - 3s_3(\tilde\chi)
+ \int_0^{\tilde\chi} d\chi\, \frac6{\tilde\chi}s_3(\chi) .
\end{equation}
Inspection shows that this is equivalent to Eq.~(\ref{eq:expect-PB}) via a change of variable, $\tau=\tau_0-\tilde\chi$.

For the magnification, \refeq{magsc} predicts
\begin{eqnarray}
\d\mmu &=& 2s_3(\tilde\chi) +\tilde\chi s'_3(\tilde\chi) + 3\int_0^{\tilde\chi} (\tilde\chi-\chi)\frac\chi{\tilde\chi}s''_3(\chi)\,d\chi
\nonumber \\
&& + \frac2{\tilde\chi} \int_0^{\tilde\chi} \left[ -s_3(\chi) + \frac12\chi^2s''_3(\chi)\right]\,d\chi;
\end{eqnarray}
repeated integration by parts again reduces this to  a form equivalent to \refeq{expect-PB-mag}.

\subsection{Potential-only mode}

The only terms left in \refeq{dgsc} that we have not tested are the observer terms, $E'(o)$ and $\partial_\parallel E'(o)$. These can be tested using a potential-only mode
\be
D({\bf x},\tau)=0 {\rm~~and~~}E({\bf x},\tau) = \Upsilon(\tau) + w_i(\tau)\,x^i.
\label{eq:PP}
\ee
This mode has no metric perturbation, $D=E_{ij}=0$, and so we expect to get $\tilde\delta_g=\delta_g$. However it does have nonzero observer terms.

Trivial evaluation shows that for the ``perturbation'' \refeq{PP}, we have $\delta z = \phi = 0$. However the convergence $\kappa$ is {\em not} zero despite the vanishing metric perturbations! Instead we have
\begin{eqnarray}
\kappa &=& -\frac12 \int_0^{\tilde\chi} d\chi\,(\tilde\chi-\chi)\frac\chi{\tilde\chi}\nabla_\perp^2 (\Upsilon''+w''_i\hat n^i\chi)
\nonumber \\
&=& \int_0^{\tilde\chi} d\chi\,\left( 1 - \frac\chi{\tilde\chi} \right) w''_i\hat n^i,
\end{eqnarray}
where in the second line the action of $\nabla_\perp^2$ is to eliminate the monopole and extract a factor of $-2\chi^{-2}$ from the dipole. Now since $w_i$ is a function only of $\tau$ we have $w''_i=d^2w_i/d\chi^2$. Double integration by parts then gives
\be
\kappa = \hat n^i w'_i(\tau_0) - \hat n^i \frac{ w_i(\tau_0) - w_i(\tau_0-\tilde\chi) }{\tilde\chi}.
\ee
Similarly we find
\begin{eqnarray}
\int_0^{\tilde\chi} d\chi \, E'' &=& \int_0^{\tilde\chi} d\chi\,(\Upsilon'' + \hat n^i w''_i \chi)
\nonumber \\
&=& \Upsilon'(\tau_0) - \Upsilon'(\tau_0-\tilde\chi) - \tilde\chi \hat n^i w'_i(\tau_0-\tilde\chi)
\nonumber \\ &&
  + \hat n^i w_i(\tau_0) - \hat n^i w_i(\tau_0-\tilde\chi).
\end{eqnarray}
Finally the remaining terms are
\be
E'-E'(o) = \Upsilon'(\tau_0-\tilde\chi) + \tilde\chi\hat n^i w'_i(\tau_0-\tilde\chi) - \Upsilon'(\tau_0).
\ee
and
\be
\partial_\parallel E'(o) = \hat n^i w'_i(\tau_0).
\ee
Assembling these pieces of \refeq{dgsc} then leads to a mass cancellation that recovers $\tilde\delta_g=\delta_g$, as expected.

The magnification equation, \refeq{magsc}, has the same nonzero pieces and a similar mass cancellation occurs, leaving the correct result $\d\mmu=0$.

%%%%%%%%%%%%%%%%%%%%%%%%%%%%%%%%%%%%%%%%%%%%%%%%%%%%%%%%%%%%%%%%%%%%%%%%%
\section{Connection with results in the literature}
\label{app:comp}

% % % % % %  % % % % % %% % % % % % % % % % % % % % % % % % %% % % % % % %% 
\subsection{Yoo et al.}

In this section, we compare our result, \refeq{nobssc}, 
\ba
\tilde{\delta}_g(\tilde\vx)
=\:& 
\d_g 
- (1+\zt)\frac{d\ln (a^3\bar n_g)}{dz}\Big|_{\!\zt} \d z +
  2 D - E_{\parallel} \vs
&- \left(1 - \frac{1+\zt}{H}\frac{d H}{d\zt}
\right) \d z - \frac{1+\zt}{H(\zt)} (D' + E'_{\parallel})\Big|_{\chit} \vs
& + 2\frac{\D x_\parallel}{\chit} - 2\hat\k 
\ea
to Eq.~(36) of \citet{yoo/etal:2009}, restricted to synchronous-comoving 
gauge:
\ba
\d_{\rm obs} =\:&  b\: (\d_m - 3 \d z) + 2D + E_{ij}\nhatt^i\nhatt^j \vs
& 
- (1+\zt) \frac{\partial}{\partial \zt} \d z 
- 2 \frac{1+\zt}{H r} \d z - \d z \vs
& 
- 5 p \d D_L - 2 \hat\k
+ \frac{1+\zt}{H}\frac{dH}{dz}\d z 
+ 2 \frac{\d r}{r}
.
\label{eq:yoo1}
\ea
We can convert their result to our notation by noting that $r = \chit$
and
\be
\D x_\parallel = \d r - \frac{1+\zt}{H(\zt)} \d z.
\ee
Note that the $\d r$ defined in Eq.~(16) of \citet{yoo/etal:2009} has
a different sign of $E_{\parallel}$ compared to ours, as discussed in \refsec{ng}.  
Further, $\partial \d z/\partial z = 1/H \partial \d z/\partial \chit$,
and using \refeq{deltaz}, \refeq{yoo1} becomes
\ba
\d_{\rm obs} =\:&  b\: (\d_m - 3 \d z) + 2D + E_{\parallel} \vs
& - \left(1 - \frac{1+\zt}{H}\frac{dH}{dz}\right)\d z 
- \frac{1+\zt}{H(\zt)} (D' + E'_{\parallel})\Big|_{\chit} \vs
& + 2 \frac{\D x_\parallel}{\chit} - 5 p \d D_L - 2 \hat\k.
\label{eq:yoo2}
\ea
Comparing the two expressions, we find two differences:  first, our result involves the time-dependence
of the number density of tracers, $d\ln \bar n_g/dz$, while this quantity
does not enter \refeq{yoo2} since the bias is defined in the uniform-redshift
gauge (see \refsec{bias}).  The second difference is the sign of the
$E_{\parallel}-$term, which goes back to the difference in sign in $\d r$
[our \refeq{Dx32}, and Eq.~(16) in \citet{yoo/etal:2009}].  This was discussed
in \refsec{ng}.  

% % % % % %  % % % % % %% % % % % % % % % % % % % % % % % % %% % % % % % %% 
\subsection{Challinor \& Lewis and Bonvin \& Durrer}

We now transform our \refeq{dgsc}
into variables in conformal Newtonian gauge in order to compare our results with
\cite{challinor/lewis:2011} and \cite{bonvin/durrer:2011}.
By using \refeq{GTscalar}, \refeq{GTvector}, and \refeq{GTmetric},
we first find the transformation law from synchronous-comoving gauge 
to conformal Newtonian gauge as
\ba
\Psi = \psi^\mathrm{(CL)} =\check\alpha =\:& -E''-\cH E' ,\nonumber\\
\Phi = \phi^\mathrm{(CL)} =-\check\varphi =\:& - \phi + \cH E' ,\nonumber\\
\check{v} =\:& E' ,\nonumber\\
\widecheck{\delta z} =\:& \delta z + \cH E' ,\nonumber\\
\check{\delta}_m =\:& \delta_m + 3\cH E' , {\rm~and}\nonumber\\
\check{\delta}_g =\:& \delta_g - \bt \cH E',
\label{eq:to_CL}
\ea
where the $\,\check{}\,$ symbol denotes the quantities in the conformal Newtonian gauge [\refeq{metric_greek}], and 
the superscript (CL) denotes the metric perturbation variable defined in 
\cite{challinor/lewis:2011}. \citet{bonvin/durrer:2011} use the Bardeen 
potentials $\Psi$, $\Phi$ as defined in the above equations.
Note that we use $\cH\equiv aH$ in order to facilitate the comparison.
We now transform the terms in \refeq{dgsc} as follows:
\ba
\delta_g + \bt \d z 
=&
\check{\delta}_g + \bt\widecheck{\d z} 
\vs
\frac{1+z}{H(z)}\partial_{\parallel}^2E'
=&
\frac{1}{\mathcal{H}}\partial_{\parallel}^2\check{v}
\vs
\left(
1 - \frac{1+z}{H}\frac{dH}{dz} 
+ \frac{2}{\chit}\frac{1+z}{H(z)}
\right)\delta z
=&
\left(
\frac{\dot{\mathcal{H}}}{\mathcal{H}^2}
+
\frac{2}{\chit\mathcal{H}}
\right)
(\widecheck{\delta z} - \cH \check{v} )
\vs
2\phi =& 
2(\check{\varphi}+\cH\check{v})
\vs
\frac{2}{\chit}E' 
=& \frac{2}{\chit}\check{v}
\vs
\phi+E'' 
=&
-\left(\check{\alpha}-\check{\varphi}\right). 
\ea
By using Einstein's equations, i.e. $\phi'=0$ [\refeq{etaprime}], and
the definition of $\check{\kappa}$
\be
\check\kappa =
-\frac{1}{2}
\int_0^{\chit}d\chi
\left(\chit-\chi\right)\frac{\chi}{\chit}\nabla^2_\perp
\left(\check\alpha -\check\varphi\right) = \k,
\ee
which is the same as the convergence in conformal Newtonian gauge
($\k$ defined in \refapp{kappa}),
\refeq{dgsc} becomes
\ba
\tilde{\delta}_g(\tilde\vx)
=\:&
\check{\delta}_g + \bt\widecheck{\d z} 
- \frac{1}{\cH}\partial^2_\parallel \hat{v}
- \left(
\frac{\dot\cH}{\cH^2}
+
\frac{2}{\chit\cH}
\right) \left[\widecheck{\d z} - \cH\check v \right]
\vs
&+2 \left[\check{\varphi} + \cH\check v\right]
-
\frac{2}{\chit}
\check v
-
\frac{2}{\chit}
\int_0^{\tilde\chi}d\chi
\left(\check\alpha -\check\varphi\right)
-2\check\k
\vs
=\:&
\check{\delta}_g + \bt\widecheck{\d z} 
- \frac{1}{\cH}\partial^2_\parallel \hat{v}
- \left(
\frac{\dot\cH}{\cH^2}
+
\frac{2}{\chit\cH}
\right) \widecheck{\d z} 
\vs
&+2 \check{\varphi} + 3 \cH\check {v}
+\check\alpha  - \frac{\dot{\check\varphi}}{\cH}
-
\frac{2}{\chit}
\int_0^{\tilde\chi}d\chi
\left(\check\alpha -\check\varphi\right)
-2\check\k,
\label{eq:CL_middle_comp}
\ea
where in the second equality, we used the identity
\be
\frac{\dot{\cH}}{\cH}\check{v}
=
\check\alpha + \cH \check{v} - \frac{\dot{\check\varphi}}{\cH},
\ee
and we drop quantities at the observer's position as is done in
\cite{challinor/lewis:2011}.

The redshift perturbation $\widecheck{\delta z}$ in conformal Newtonian gauge
can be calculated from \refeq{deltaz_sc} and the gauge transformation 
\refeq{to_CL} as 
\be 
\widecheck{\d z}
=
\partial_\parallel \check{v}
-
\check\alpha 
-
\int_0^{\chit}d\chi \left(\check\a -\check\varphi\right)'.
\ee
Rewriting the equation in terms of the variables in 
\citet{challinor/lewis:2011}
($\check\a = \psi$,
$\check\varphi = -\phi$,
and
$\partial_\parallel \check{v} = \bf{v}\cdot\hat{\bf{n}}$),
we find the redshift perturbation as
\be
\widecheck{\d z}
=
{\bf{v}}\cdot\hat{\bf{n}}
- \psi^{\rm (CL)}
- \int_0^{\chit} d\chi \left(\psi^{\rm (CL)} +\phi^{\rm (CL)}\right)'.
\ee
A similar change of variables can be made for the observed galaxy density
contrast in \refeq{CL_middle_comp}, which yields
\ba
\tilde{\delta}_g(\tilde\vx)
=\:&
\Delta_n(\hat{\mathbf n},z)
+ 3\cH \check{v}
\label{eq:final_comp}
\ea
where $\Delta_n(\hat{\bf{n}},z)$ is defined in Eq.~(30) of
\citet{challinor/lewis:2011}.  The additional term of $3\cH\check{v}$ is due to 
the fact that the overdensity $\Delta_n$ 
is defined with respect to the physical, rather than comoving galaxy density.  
In performing the gauge transformation $t\rightarrow \check t = t + T$,
we thus obtain an addition term from 
$\check{\ln a^3} = \ln a^3 + 3 a H T = \ln a^3 + 3 \cH \check v$.  

To compare the magnification terms, we transform \refeq{magsc}
to the conformal Newtonian gauge, yielding
\ba
\d\mmu
=& 
-2(\check\varphi+\cH\check{v})
+
\frac{2}{\chit} \check{v}
+
2\check{\kappa}
-
\frac{2}{\chit}\int_0^{\chit}d\chi
\left(\check\a-\check\varphi\right)
\vs
&-2
\left(
1-\frac{1}{\cH\chit}
\right)\left(\widecheck{\d z} - \cH \check{v}\right)
\vs
=& 
2\kappa
+2\phi^{\rm (CL)}
-
\frac{2}{\chit}\int_0^{\chit}d\chi
\left(\phi^{\rm(CL)}+\psi^{\rm(CL)}\right)
\vs
&+2\left(
\frac{1}{\cH\chit}-1
\right)
\widecheck{\d z},
\label{eq:dmmHere}
\ea
which has to be compared to all terms $\propto 5s$
in Eq.~(37) of \cite{challinor/lewis:2011}: 
\ba
&\k - \frac{1}{\chit} \int d\chi (\phi+\psi) 
+ \phi \vs
&+ \left(\frac{1}{\cH \chit} - 1\right)
\left[- \psi - \int d\chi (\phi'+\psi') + \hat n \cdot v\right]
\label{eq:dmmCL}
\ea
As $5s$ is the same as $2\Q$ in our notation, our formula for
$\d\mmu$ is also in agreement with \cite{challinor/lewis:2011}.  

We can also write \refeq{CL_middle_comp} in terms of the variables in \citet{bonvin/durrer:2011},
\ba
\check{\d}_g+\bt\widecheck{\d}z =& D_s, \nonumber \\
\check\a =& \Psi, \nonumber \\
\check\varphi =& -\Phi, \nonumber \\
\partial_\parallel \check{v} =& -\bf{V}\cdot\hat{\bf{n}}, {\rm~~and}\nonumber \\
\partial^2_\parallel \check{v} =& \partial_r\left(\bf{V}\cdot\hat{\bf{n}}\right),
\ea
to arrive at the same relation \refeq{final_comp}, after identifying
\ba
\kappa
=
\frac{1}{2r_S}\int_0^{r_S}
dr
\left[
\frac{r_S-r}{r}\Delta_\Omega
\right]
\left(\Phi+\Psi\right)
\ea
and $\Delta_n = \Delta$ as defined in Ref.~\cite{bonvin/durrer:2011}.  

%%%%%%%%%%%%%%%%%%%%%%%%%%%%%%%%%%%%%%%%%%%%%%%%%%%%%%%%%%%%%%%%%%%%%%%%%%%%%%%%
%\bibliography{gaugePk}

\begin{thebibliography}{32}
\expandafter\ifx\csname natexlab\endcsname\relax\def\natexlab#1{#1}\fi
\expandafter\ifx\csname bibnamefont\endcsname\relax
  \def\bibnamefont#1{#1}\fi
\expandafter\ifx\csname bibfnamefont\endcsname\relax
  \def\bibfnamefont#1{#1}\fi
\expandafter\ifx\csname citenamefont\endcsname\relax
  \def\citenamefont#1{#1}\fi
\expandafter\ifx\csname url\endcsname\relax
  \def\url#1{\texttt{#1}}\fi
\expandafter\ifx\csname urlprefix\endcsname\relax\def\urlprefix{URL }\fi
\providecommand{\bibinfo}[2]{#2}
\providecommand{\eprint}[2][]{\url{#2}}

\bibitem[{\citenamefont{{Dalal} et~al.}(2008)\citenamefont{{Dalal}, {Dor{\'e}},
  {Huterer}, and {Shirokov}}}]{DalalEtal08}
\bibinfo{author}{\bibfnamefont{N.}~\bibnamefont{{Dalal}}},
  \bibinfo{author}{\bibfnamefont{O.}~\bibnamefont{{Dor{\'e}}}},
  \bibinfo{author}{\bibfnamefont{D.}~\bibnamefont{{Huterer}}},
  \bibnamefont{and}
  \bibinfo{author}{\bibfnamefont{A.}~\bibnamefont{{Shirokov}}},
  \bibinfo{journal}{\prd} \textbf{\bibinfo{volume}{77}},
  \bibinfo{pages}{123514} (\bibinfo{year}{2008}), \eprint{0710.4560}.

\bibitem[{\citenamefont{{Slosar} et~al.}(2008)\citenamefont{{Slosar}, {Hirata},
  {Seljak}, {Ho}, and {Padmanabhan}}}]{SlosarEtal}
\bibinfo{author}{\bibfnamefont{A.}~\bibnamefont{{Slosar}}},
  \bibinfo{author}{\bibfnamefont{C.}~\bibnamefont{{Hirata}}},
  \bibinfo{author}{\bibfnamefont{U.}~\bibnamefont{{Seljak}}},
  \bibinfo{author}{\bibfnamefont{S.}~\bibnamefont{{Ho}}}, \bibnamefont{and}
  \bibinfo{author}{\bibfnamefont{N.}~\bibnamefont{{Padmanabhan}}},
  \bibinfo{journal}{\jcap} \textbf{\bibinfo{volume}{8}}, \bibinfo{pages}{31}
  (\bibinfo{year}{2008}), \eprint{0805.3580}.

\bibitem[{\citenamefont{{Eisenstein} et~al.}(2011)\citenamefont{{Eisenstein},
  {Weinberg}, {Agol}, {Aihara}, {Allende Prieto}, {Anderson}, {Arns},
  {Aubourg}, {Bailey}, {Balbinot} et~al.}}]{BOSS}
\bibinfo{author}{\bibfnamefont{D.~J.} \bibnamefont{{Eisenstein}}},
  \bibinfo{author}{\bibfnamefont{D.~H.} \bibnamefont{{Weinberg}}},
  \bibinfo{author}{\bibfnamefont{E.}~\bibnamefont{{Agol}}},
  \bibinfo{author}{\bibfnamefont{H.}~\bibnamefont{{Aihara}}},
  \bibinfo{author}{\bibfnamefont{C.}~\bibnamefont{{Allende Prieto}}},
  \bibinfo{author}{\bibfnamefont{S.~F.} \bibnamefont{{Anderson}}},
  \bibinfo{author}{\bibfnamefont{J.~A.} \bibnamefont{{Arns}}},
  \bibinfo{author}{\bibfnamefont{E.}~\bibnamefont{{Aubourg}}},
  \bibinfo{author}{\bibfnamefont{S.}~\bibnamefont{{Bailey}}},
  \bibinfo{author}{\bibfnamefont{E.}~\bibnamefont{{Balbinot}}},
  \bibnamefont{et~al.}, \bibinfo{journal}{ArXiv e-prints}
  (\bibinfo{year}{2011}), \eprint{1101.1529}.

\bibitem[{\citenamefont{{Hill} et~al.}(2008)\citenamefont{{Hill}, {Gebhardt},
  {Komatsu}, {Drory}, {MacQueen}, {Adams}, {Blanc}, {Koehler}, {Rafal}, {Roth}
  et~al.}}]{hetdex}
\bibinfo{author}{\bibfnamefont{G.~J.} \bibnamefont{{Hill}}},
  \bibinfo{author}{\bibfnamefont{K.}~\bibnamefont{{Gebhardt}}},
  \bibinfo{author}{\bibfnamefont{E.}~\bibnamefont{{Komatsu}}},
  \bibinfo{author}{\bibfnamefont{N.}~\bibnamefont{{Drory}}},
  \bibinfo{author}{\bibfnamefont{P.~J.} \bibnamefont{{MacQueen}}},
  \bibinfo{author}{\bibfnamefont{J.}~\bibnamefont{{Adams}}},
  \bibinfo{author}{\bibfnamefont{G.~A.} \bibnamefont{{Blanc}}},
  \bibinfo{author}{\bibfnamefont{R.}~\bibnamefont{{Koehler}}},
  \bibinfo{author}{\bibfnamefont{M.}~\bibnamefont{{Rafal}}},
  \bibinfo{author}{\bibfnamefont{M.~M.} \bibnamefont{{Roth}}},
  \bibnamefont{et~al.}, in \emph{\bibinfo{booktitle}{Panoramic Views of Galaxy
  Formation and Evolution}}, edited by
  \bibinfo{editor}{\bibnamefont{{T.~Kodama, T.~Yamada, \& K.~Aoki}}}
  (\bibinfo{year}{2008}), vol. \bibinfo{volume}{399} of
  \emph{\bibinfo{series}{Astronomical Society of the Pacific Conference
  Series}}, pp. \bibinfo{pages}{115--+}, \eprint{0806.0183}.

\bibitem[{\citenamefont{{Schlegel} et~al.}(2011)\citenamefont{{Schlegel},
  {Abdalla}, {Abraham}, {Ahn}, {Allende Prieto}, {Annis}, {Aubourg}, {Azzaro},
  {Baltay}, {Baugh} et~al.}}]{BigBOSS}
\bibinfo{author}{\bibfnamefont{D.}~\bibnamefont{{Schlegel}}},
  \bibinfo{author}{\bibfnamefont{F.}~\bibnamefont{{Abdalla}}},
  \bibinfo{author}{\bibfnamefont{T.}~\bibnamefont{{Abraham}}},
  \bibinfo{author}{\bibfnamefont{C.}~\bibnamefont{{Ahn}}},
  \bibinfo{author}{\bibfnamefont{C.}~\bibnamefont{{Allende Prieto}}},
  \bibinfo{author}{\bibfnamefont{J.}~\bibnamefont{{Annis}}},
  \bibinfo{author}{\bibfnamefont{E.}~\bibnamefont{{Aubourg}}},
  \bibinfo{author}{\bibfnamefont{M.}~\bibnamefont{{Azzaro}}},
  \bibinfo{author}{\bibfnamefont{S.~B.~C.} \bibnamefont{{Baltay}}},
  \bibinfo{author}{\bibfnamefont{C.}~\bibnamefont{{Baugh}}},
  \bibnamefont{et~al.}, \bibinfo{journal}{ArXiv e-prints}
  (\bibinfo{year}{2011}), \eprint{1106.1706}.

\bibitem[{\citenamefont{{Sachs} and {Wolfe}}(1967)}]{SW67}
\bibinfo{author}{\bibfnamefont{R.~K.} \bibnamefont{{Sachs}}} \bibnamefont{and}
  \bibinfo{author}{\bibfnamefont{A.~M.} \bibnamefont{{Wolfe}}},
  \bibinfo{journal}{\apj} \textbf{\bibinfo{volume}{147}}, \bibinfo{pages}{73}
  (\bibinfo{year}{1967}).

\bibitem[{\citenamefont{{Sasaki}}(1987)}]{Sasaki87}
\bibinfo{author}{\bibfnamefont{M.}~\bibnamefont{{Sasaki}}},
  \bibinfo{journal}{\mnras} \textbf{\bibinfo{volume}{228}},
  \bibinfo{pages}{653} (\bibinfo{year}{1987}).

\bibitem[{\citenamefont{{Bonvin} et~al.}(2006)\citenamefont{{Bonvin}, {Durrer},
  and {Gasparini}}}]{BonvinEtal06}
\bibinfo{author}{\bibfnamefont{C.}~\bibnamefont{{Bonvin}}},
  \bibinfo{author}{\bibfnamefont{R.}~\bibnamefont{{Durrer}}}, \bibnamefont{and}
  \bibinfo{author}{\bibfnamefont{M.~A.} \bibnamefont{{Gasparini}}},
  \bibinfo{journal}{\prd} \textbf{\bibinfo{volume}{73}},
  \bibinfo{pages}{023523} (\bibinfo{year}{2006}),
  \eprint{arXiv:astro-ph/0511183}.

\bibitem[{\citenamefont{{Yoo} et~al.}(2009)\citenamefont{{Yoo}, {Fitzpatrick},
  and {Zaldarriaga}}}]{yoo/etal:2009}
\bibinfo{author}{\bibfnamefont{J.}~\bibnamefont{{Yoo}}},
  \bibinfo{author}{\bibfnamefont{A.~L.} \bibnamefont{{Fitzpatrick}}},
  \bibnamefont{and}
  \bibinfo{author}{\bibfnamefont{M.}~\bibnamefont{{Zaldarriaga}}},
  \bibinfo{journal}{\prd} \textbf{\bibinfo{volume}{80}},
  \bibinfo{pages}{083514} (\bibinfo{year}{2009}), \eprint{0907.0707}.

\bibitem[{\citenamefont{{Yoo}}(2010)}]{yoo:2010}
\bibinfo{author}{\bibfnamefont{J.}~\bibnamefont{{Yoo}}},
  \bibinfo{journal}{\prd} \textbf{\bibinfo{volume}{82}},
  \bibinfo{pages}{083508} (\bibinfo{year}{2010}), \eprint{1009.3021}.

\bibitem[{\citenamefont{{Challinor} and {Lewis}}(2011)}]{challinor/lewis:2011}
\bibinfo{author}{\bibfnamefont{A.}~\bibnamefont{{Challinor}}} \bibnamefont{and}
  \bibinfo{author}{\bibfnamefont{A.}~\bibnamefont{{Lewis}}},
  \bibinfo{journal}{ArXiv e-prints}  (\bibinfo{year}{2011}),
  \eprint{1105.5292}.

\bibitem[{\citenamefont{{Bruni} et~al.}(2011)\citenamefont{{Bruni},
  {Crittenden}, {Koyama}, {Maartens}, {Pitrou}, and {Wands}}}]{BruniEtal}
\bibinfo{author}{\bibfnamefont{M.}~\bibnamefont{{Bruni}}},
  \bibinfo{author}{\bibfnamefont{R.}~\bibnamefont{{Crittenden}}},
  \bibinfo{author}{\bibfnamefont{K.}~\bibnamefont{{Koyama}}},
  \bibinfo{author}{\bibfnamefont{R.}~\bibnamefont{{Maartens}}},
  \bibinfo{author}{\bibfnamefont{C.}~\bibnamefont{{Pitrou}}}, \bibnamefont{and}
  \bibinfo{author}{\bibfnamefont{D.}~\bibnamefont{{Wands}}},
  \bibinfo{journal}{ArXiv e-prints}  (\bibinfo{year}{2011}),
  \eprint{1106.3999}.

\bibitem[{\citenamefont{{Bonvin} and {Durrer}}(2011)}]{bonvin/durrer:2011}
\bibinfo{author}{\bibfnamefont{C.}~\bibnamefont{{Bonvin}}} \bibnamefont{and}
  \bibinfo{author}{\bibfnamefont{R.}~\bibnamefont{{Durrer}}},
  \bibinfo{journal}{ArXiv e-prints}  (\bibinfo{year}{2011}),
  \eprint{1105.5280}.

\bibitem[{\citenamefont{{Dodelson} et~al.}(2008)\citenamefont{{Dodelson},
  {Schmidt}, and {Vallinotto}}}]{DodelsonEtal08}
\bibinfo{author}{\bibfnamefont{S.}~\bibnamefont{{Dodelson}}},
  \bibinfo{author}{\bibfnamefont{F.}~\bibnamefont{{Schmidt}}},
  \bibnamefont{and}
  \bibinfo{author}{\bibfnamefont{A.}~\bibnamefont{{Vallinotto}}},
  \bibinfo{journal}{\prd} \textbf{\bibinfo{volume}{78}},
  \bibinfo{pages}{043508} (\bibinfo{year}{2008}), \eprint{0806.0331}.

\bibitem[{\citenamefont{{Hirata}}(2009{\natexlab{a}})}]{hirataquasar}
\bibinfo{author}{\bibfnamefont{C.~M.} \bibnamefont{{Hirata}}},
  \bibinfo{journal}{\jcap} \textbf{\bibinfo{volume}{9}}, \bibinfo{pages}{11}
  (\bibinfo{year}{2009}{\natexlab{a}}), \eprint{0907.0703}.

\bibitem[{\citenamefont{{Schmidt} et~al.}(2009)\citenamefont{{Schmidt}, {Rozo},
  {Dodelson}, {Hui}, and {Sheldon}}}]{schmidtetal09}
\bibinfo{author}{\bibfnamefont{F.}~\bibnamefont{{Schmidt}}},
  \bibinfo{author}{\bibfnamefont{E.}~\bibnamefont{{Rozo}}},
  \bibinfo{author}{\bibfnamefont{S.}~\bibnamefont{{Dodelson}}},
  \bibinfo{author}{\bibfnamefont{L.}~\bibnamefont{{Hui}}}, \bibnamefont{and}
  \bibinfo{author}{\bibfnamefont{E.}~\bibnamefont{{Sheldon}}},
  \bibinfo{journal}{Physical Review Letters} \textbf{\bibinfo{volume}{103}},
  \bibinfo{pages}{051301} (\bibinfo{year}{2009}), \eprint{0904.4702}.

\bibitem[{\citenamefont{{McDonald} and {Roy}}(2009)}]{McDonaldRoy}
\bibinfo{author}{\bibfnamefont{P.}~\bibnamefont{{McDonald}}} \bibnamefont{and}
  \bibinfo{author}{\bibfnamefont{A.}~\bibnamefont{{Roy}}},
  \bibinfo{journal}{\jcap} \textbf{\bibinfo{volume}{8}}, \bibinfo{pages}{20}
  (\bibinfo{year}{2009}), \eprint{0902.0991}.

\bibitem[{\citenamefont{{Hirata}}(2009{\natexlab{b}})}]{Hirata09}
\bibinfo{author}{\bibfnamefont{C.~M.} \bibnamefont{{Hirata}}},
  \bibinfo{journal}{\mnras} \textbf{\bibinfo{volume}{399}},
  \bibinfo{pages}{1074} (\bibinfo{year}{2009}{\natexlab{b}}),
  \eprint{0903.4929}.

\bibitem[{\citenamefont{{Kaiser}}(1984)}]{Kaiser84}
\bibinfo{author}{\bibfnamefont{N.}~\bibnamefont{{Kaiser}}},
  \bibinfo{journal}{\apjl} \textbf{\bibinfo{volume}{284}}, \bibinfo{pages}{L9}
  (\bibinfo{year}{1984}).

\bibitem[{\citenamefont{{Mo} and {White}}(1996)}]{MoWhite}
\bibinfo{author}{\bibfnamefont{H.~J.} \bibnamefont{{Mo}}} \bibnamefont{and}
  \bibinfo{author}{\bibfnamefont{S.~D.~M.} \bibnamefont{{White}}},
  \bibinfo{journal}{\mnras} \textbf{\bibinfo{volume}{282}},
  \bibinfo{pages}{347} (\bibinfo{year}{1996}), \eprint{arXiv:astro-ph/9512127}.

\bibitem[{\citenamefont{{Reid} et~al.}(2010)\citenamefont{{Reid}, {Verde},
  {Dolag}, {Matarrese}, and {Moscardini}}}]{ReidEtal}
\bibinfo{author}{\bibfnamefont{B.~A.} \bibnamefont{{Reid}}},
  \bibinfo{author}{\bibfnamefont{L.}~\bibnamefont{{Verde}}},
  \bibinfo{author}{\bibfnamefont{K.}~\bibnamefont{{Dolag}}},
  \bibinfo{author}{\bibfnamefont{S.}~\bibnamefont{{Matarrese}}},
  \bibnamefont{and}
  \bibinfo{author}{\bibfnamefont{L.}~\bibnamefont{{Moscardini}}},
  \bibinfo{journal}{\jcap} \textbf{\bibinfo{volume}{7}}, \bibinfo{pages}{13}
  (\bibinfo{year}{2010}), \eprint{1004.1637}.

\bibitem[{\citenamefont{{Kaiser}}(1987)}]{kaiser:1987}
\bibinfo{author}{\bibfnamefont{N.}~\bibnamefont{{Kaiser}}},
  \bibinfo{journal}{\mnras} \textbf{\bibinfo{volume}{227}}, \bibinfo{pages}{1}
  (\bibinfo{year}{1987}).

\bibitem[{\citenamefont{{Chisari} and
  {Zaldarriaga}}(2011)}]{ChisariZaldarriaga}
\bibinfo{author}{\bibfnamefont{N.~E.} \bibnamefont{{Chisari}}}
  \bibnamefont{and}
  \bibinfo{author}{\bibfnamefont{M.}~\bibnamefont{{Zaldarriaga}}},
  \bibinfo{journal}{ArXiv e-prints}  (\bibinfo{year}{2011}),
  \eprint{1101.3555}.

\bibitem[{\citenamefont{{Komatsu} et~al.}(2011)\citenamefont{{Komatsu},
  {Smith}, {Dunkley}, {Bennett}, {Gold}, {Hinshaw}, {Jarosik}, {Larson},
  {Nolta}, {Page} et~al.}}]{wmap7}
\bibinfo{author}{\bibfnamefont{E.}~\bibnamefont{{Komatsu}}},
  \bibinfo{author}{\bibfnamefont{K.~M.} \bibnamefont{{Smith}}},
  \bibinfo{author}{\bibfnamefont{J.}~\bibnamefont{{Dunkley}}},
  \bibinfo{author}{\bibfnamefont{C.~L.} \bibnamefont{{Bennett}}},
  \bibinfo{author}{\bibfnamefont{B.}~\bibnamefont{{Gold}}},
  \bibinfo{author}{\bibfnamefont{G.}~\bibnamefont{{Hinshaw}}},
  \bibinfo{author}{\bibfnamefont{N.}~\bibnamefont{{Jarosik}}},
  \bibinfo{author}{\bibfnamefont{D.}~\bibnamefont{{Larson}}},
  \bibinfo{author}{\bibfnamefont{M.~R.} \bibnamefont{{Nolta}}},
  \bibinfo{author}{\bibfnamefont{L.}~\bibnamefont{{Page}}},
  \bibnamefont{et~al.}, \bibinfo{journal}{\apjs}
  \textbf{\bibinfo{volume}{192}}, \bibinfo{pages}{18} (\bibinfo{year}{2011}),
  \eprint{1001.4538}.

\bibitem[{\citenamefont{{Schmidt}}(2010)}]{Schmidt10}
\bibinfo{author}{\bibfnamefont{F.}~\bibnamefont{{Schmidt}}},
  \bibinfo{journal}{\prd} \textbf{\bibinfo{volume}{82}},
  \bibinfo{pages}{063001} (\bibinfo{year}{2010}), \eprint{1005.4063}.

\bibitem[{\citenamefont{{Baldauf} et~al.}(2011)\citenamefont{{Baldauf},
  {Seljak}, {Senatore}, and {Zaldarriaga}}}]{BaldaufEtal}
\bibinfo{author}{\bibfnamefont{T.}~\bibnamefont{{Baldauf}}},
  \bibinfo{author}{\bibfnamefont{U.}~\bibnamefont{{Seljak}}},
  \bibinfo{author}{\bibfnamefont{L.}~\bibnamefont{{Senatore}}},
  \bibnamefont{and}
  \bibinfo{author}{\bibfnamefont{M.}~\bibnamefont{{Zaldarriaga}}},
  \bibinfo{journal}{ArXiv e-prints}  (\bibinfo{year}{2011}),
  \eprint{1106.5507}.

\bibitem[{\citenamefont{{Desjacques} et~al.}(2011)\citenamefont{{Desjacques},
  {Jeong}, and {Schmidt}}}]{DJS}
\bibinfo{author}{\bibfnamefont{V.}~\bibnamefont{{Desjacques}}},
  \bibinfo{author}{\bibfnamefont{D.}~\bibnamefont{{Jeong}}}, \bibnamefont{and}
  \bibinfo{author}{\bibfnamefont{F.}~\bibnamefont{{Schmidt}}},
  \bibinfo{journal}{ArXiv e-prints}  (\bibinfo{year}{2011}),
  \eprint{1105.3628}.

\bibitem[{\citenamefont{{Schmidt} and {Kamionkowski}}(2010)}]{fsmk}
\bibinfo{author}{\bibfnamefont{F.}~\bibnamefont{{Schmidt}}} \bibnamefont{and}
  \bibinfo{author}{\bibfnamefont{M.}~\bibnamefont{{Kamionkowski}}},
  \bibinfo{journal}{\prd} \textbf{\bibinfo{volume}{82}},
  \bibinfo{pages}{103002} (\bibinfo{year}{2010}), \eprint{1008.0638}.

\bibitem[{\citenamefont{{Wands} and {Slosar}}(2009)}]{WandsSlosar}
\bibinfo{author}{\bibfnamefont{D.}~\bibnamefont{{Wands}}} \bibnamefont{and}
  \bibinfo{author}{\bibfnamefont{A.}~\bibnamefont{{Slosar}}},
  \bibinfo{journal}{\prd} \textbf{\bibinfo{volume}{79}},
  \bibinfo{pages}{123507} (\bibinfo{year}{2009}), \eprint{0902.1084}.

\bibitem[{\citenamefont{{Bardeen}}(1980)}]{bardeen:1980}
\bibinfo{author}{\bibfnamefont{J.~M.} \bibnamefont{{Bardeen}}},
  \bibinfo{journal}{\prd} \textbf{\bibinfo{volume}{22}}, \bibinfo{pages}{1882}
  (\bibinfo{year}{1980}).

\bibitem[{\citenamefont{{Ma} and {Bertschinger}}(1995)}]{ma/bertschinger:1995}
\bibinfo{author}{\bibfnamefont{C.}~\bibnamefont{{Ma}}} \bibnamefont{and}
  \bibinfo{author}{\bibfnamefont{E.}~\bibnamefont{{Bertschinger}}},
  \bibinfo{journal}{\apj} \textbf{\bibinfo{volume}{455}}, \bibinfo{pages}{7}
  (\bibinfo{year}{1995}), \eprint{arXiv:astro-ph/9506072}.

\bibitem[{\citenamefont{Hu and Cooray}(2000)}]{HuCooray00}
\bibinfo{author}{\bibfnamefont{W.}~\bibnamefont{Hu}} \bibnamefont{and}
  \bibinfo{author}{\bibfnamefont{A.}~\bibnamefont{Cooray}},
  \bibinfo{journal}{Phys. Rev. D} \textbf{\bibinfo{volume}{63}},
  \bibinfo{pages}{023504} (\bibinfo{year}{2000}).

\end{thebibliography}

\end{document}